\documentclass[11pt]{article}

\usepackage[margin=1in]{geometry}
\usepackage{amsmath}
\usepackage{amssymb}
\usepackage{booktabs}
\usepackage{array}
\usepackage{graphicx}
\usepackage{microtype}
\usepackage{xcolor}
\usepackage{tikz}
\usetikzlibrary{arrows.meta, calc, positioning}
\usepackage[section]{placeins}

\usepackage{textcomp}
\usepackage[colorlinks=true,linkcolor=blue!50!black,citecolor=blue!50!black,urlcolor=blue!50!black]{hyperref}

\newcommand{\artifact}[1]{\hspace{0pt}{\sloppy\texttt{\small #1}}\hspace{0pt}}

\newcommand{\us}{\ensuremath{\,\mathrm{us}}}

\title{Memory-Bound but Not Bandwidth-Limited: The Physical AI Inference Gap in Batch-1 LLM Decode \\ \large A 44-Cell Cross-GPU Study of Memory Floors, CUDA Graphs and Quantized Decode}

\author{Josef Chen \\ KAIKAKU \\ \texttt{josef@kaikaku.ai}}

\date{May 2026}

\begin{document}

\maketitle

\begin{abstract}
Physical AI systems, including robots, autonomous vehicles, embodied agents
and edge copilots, often run a different inference workload from cloud LLM
serving: single-stream, batch-1 autoregressive decode, where one robot,
camera feed or user session waits on the next token. This workload is usually
described as memory-bandwidth-bound. Each decode step streams model weights
and the active KV cache, so latency should scale with peak HBM bandwidth. We
show that this account is true but incomplete.

We measure batch-1 decode for three 7 to 8B-class GQA transformers across
four NVIDIA GPUs: H100 SXM5, A100-80GB SXM4, L40S and L4. We evaluate context
lengths from 2048 to 16384, producing 44 valid cells under a controlled bf16
SDPA setup. The achieved fraction of peak HBM bandwidth falls as peak
bandwidth rises. On the headline Qwen-2.5-7B ctx${=}2048$ cell, an L4 reaches
roughly 81\% of its analytic memory floor, while an H100 reaches only 27\%.
Physical-AI decode is memory-dominated, but faster memory does not translate
into proportional latency gains.

We test the missing term with a CUDA Graphs A/B experiment. On H100 at
ctx${=}2048$, CUDA Graphs improves decode latency by $1.259\times$ across
$N{=}10$ fresh sessions, with a 95\% bootstrap confidence interval of
$[1.253, 1.267]$. On L4, the same intervention gives only $1.028\times$.
This isolates a launch-side overhead that becomes visible on fast GPUs but
remains mostly hidden on slower, bandwidth-bound GPUs.

The deployment implication is that memory savings matter only when the
runtime realises them. On L4, bf16 decode sits close to the memory floor,
but common quantised paths do not recover the expected $4\times$
weight-traffic reduction: bnb-nf4 reaches 59.36\,ms/step and
AutoAWQ\,+\,Marlin reaches 45.24\,ms/step from a 62.32\,ms bf16 baseline.
GPTQ\,+\,ExLlamaV2, with Ada-tuned int4 kernels, reaches 17.36\,ms/step. For
physical-AI inference, the bottleneck is not peak bandwidth alone, but the
interaction between memory traffic, launch overhead and runtime support for
compressed weight movement.

\end{abstract}

\section{Introduction}
\label{sec:intro}

Physical-AI inference, robotics policy heads, autonomous-driving language
copilots and on-device assistants share a workload that LLM-serving research
mostly does not target: single-stream, batch-1, autoregressive decode of
7--8B-class transformers. The user is one human, one robot or one camera
feed; latency between tokens is the user-facing metric; throughput-oriented
batching is not available because there is no batch.

The standard account of batch-1 decode is that it is HBM-bandwidth-bound:
each decode step streams the model weights and the per-layer KV cache through
global memory once, arithmetic intensity is too low to keep the SMs busy, and
step time is approximately $(W + K)/B_{\mathrm{peak}}$. This account motivates
a sizeable fraction of recent serving work: KV cache
compression~\cite{zhang2023h2o,li2024snapkv,xiao2024streamingllm,liu2024kivi,hooper2024kvquant},
weight quantisation~\cite{frantar2022gptq,lin2024awq,xiao2022smoothquant} and KV offloading.
It also drives the implicit assumption that
deployment ladder L4\,$\rightarrow$\,L40S\,$\rightarrow$\,A100\,$\rightarrow$\,H100
is the cost-per-token ladder, because each rung adds peak HBM bandwidth.

This paper is a measurement study, not a modeling paper. We measure decode
step time across four NVIDIA GPUs that span more than a $10\times$ range in
peak HBM bandwidth (L4 at 300\,GB/s~\cite{nvidia_l4_ds} through H100 SXM5
at 3350\,GB/s~\cite{nvidia_h100_ds}), three 7--8B-class GQA
architectures~\cite{ainslie2023gqa} (Qwen-2.5-7B~\cite{qwen2025qwen25},
Mistral-7B-v0.3~\cite{jiang2023mistral}, Llama-3.1-8B), and four context
lengths between 2048 and 16384, all at batch 1 with bf16 weights and sdpa
attention on Modal cloud hosts~\cite{modal2025pricing}. The directly-measured observed-over-floor ratio
$R_{\mathrm{floor}} = t_{\mathrm{obs}}/t_{\mathrm{floor}}$, with
$t_{\mathrm{floor}} = (W+K)/B_{\mathrm{peak}}$, falls from $\sim 0.8$ on L4
to $\sim 0.25$ on H100 at short contexts. Translated into achieved bandwidth,
the L4 reaches roughly 81\% of its peak HBM bandwidth on this workload while
the H100 reaches only 27\%. The HBM-bound prediction therefore overestimates
achievable speedup on fast silicon by a factor of three on this workload class.

\paragraph{Mechanism, falsified at $N{=}10$.} The load-bearing falsifiable
claim is mechanistic: the H100 gap is per-kernel CPU launch overhead, not
silicon. We test it with a CUDA Graphs A/B that touches the launch term and
only the launch term. On the H100 ctx${=}2048$ headline cell at $N{=}10$
independent Modal sessions, the within-session graphed/eager speedup is
$1.259\times$ with 95\% bootstrap CI $[1.253, 1.267]$ and cross-session CV
0.9\%; on L4 at the same cell the same intervention yields $1.028\times$.
The prediction would have been killed by an H100 speedup under $\sim 1.15\times$
in most sessions, or by an L4 speedup over $\sim 1.15\times$ in any session.
Neither condition occurred.

\paragraph{Deployment implication.} The substantive implication is the
inverted cost-per-token ladder. With backend-pinned SDPA measurements,
PyTorch's default SDPA dispatcher ($36.05\us$/layer) outperforms explicit
\texttt{FLASH\_ATTENTION} context ($44.35\us$), FlashInfer ($48.20\us$) and
FlashAttention-3 ($79.25\us$) at H100 batch-1 single-decode shape; the kernel
choice is not the binding constraint. The kernel choice on L4 \textit{is} the
binding constraint: substituting GPTQ${+}$ExLlamaV2 for the default Marlin
path cuts Qwen-2.5-7B step time from 62.32\,ms (bf16) to 17.36\,ms, a
$3.59\times$ speedup that AutoAWQ${+}$Marlin (45.24\,ms) does not deliver on
Ada SM89. An L4 with ExLlamaV2 runs Qwen-2.5-7B at 17.36\,ms/step; an H100
with CUDA Graphs runs the same workload at 11.78\,ms/step. For the streaming
workloads that dominate physical AI, the GPU upgrade closes less of the
gap than the kernel choice on the cheaper silicon does.

\begin{figure}[!ht]
\centering
\includegraphics[width=0.95\linewidth]{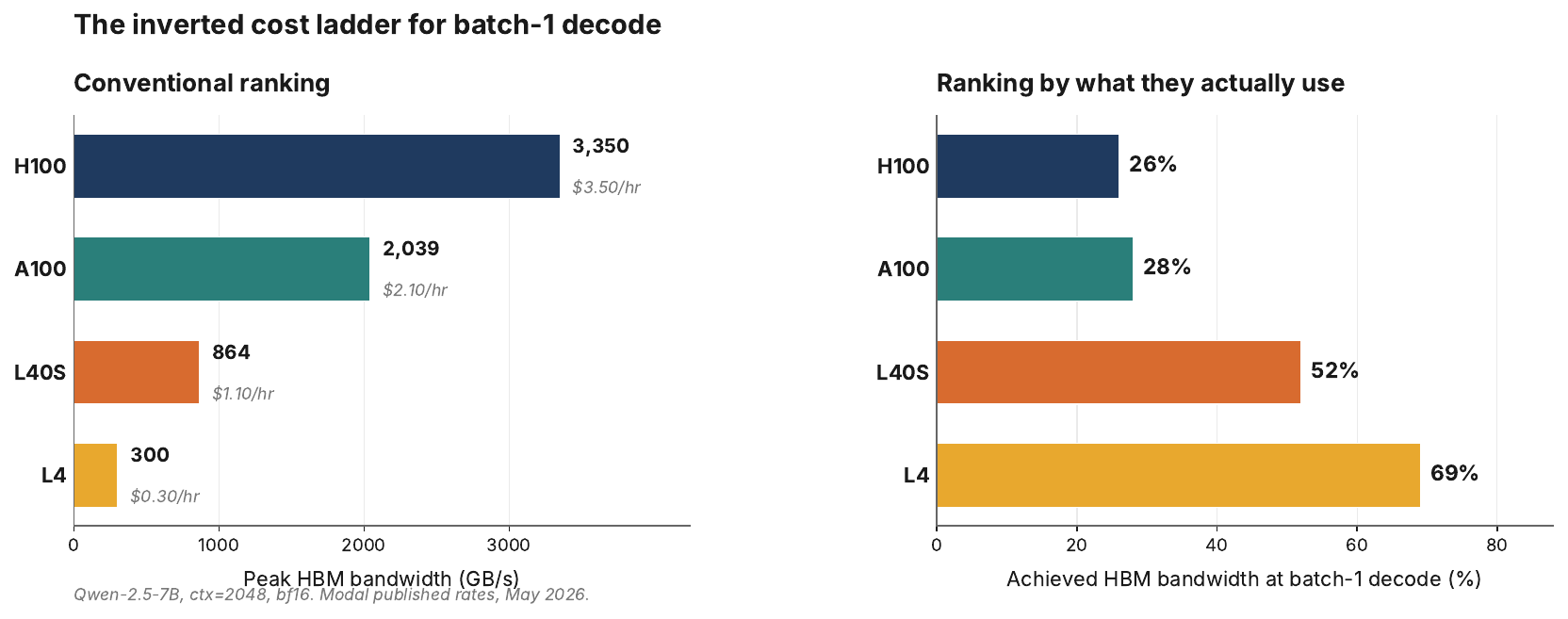}
\caption{The inverted deployment ladder for single-stream 7--8B decode.
Left: achieved fraction of peak HBM bandwidth at Qwen-2.5-7B ctx${=}2048$
batch~1 across four NVIDIA GPUs; L4 reaches 79\% of its peak, H100 only
27\%. Right: step time for the same workload after applying the best
lever per GPU; L4 with GPTQ${+}$ExLlamaV2 (17.36\,ms) closes most of the
gap to H100 with CUDA Graphs (11.78\,ms) despite the H100 having
$11\times$ the peak HBM bandwidth.}
\label{fig:hero}
\end{figure}

\paragraph{Contributions.}
\begin{enumerate}
  \item A measurement study of single-stream decode at batch 1 across three
        7--8B-class GQA architectures, four NVIDIA GPUs and four context
        lengths, with controlled software stack, warmup and 30 measured
        steps per cell (44 valid cells, 4 OOMs on L4 at long context, all
        diagnosed as A/B-protocol artifacts not capacity ceilings).
  \item A CUDA Graphs A/B falsification test at $N{=}10$ within-session and
        cross-session, confirming the launch-tax interpretation on H100
        short-context decode and rejecting it on L4 with a null result.
  \item Backend-pinned attention-kernel measurements at H100 batch-1
        single-decode shape, showing that PyTorch's default SDPA dispatcher
        beats every explicit backend including FlashAttention-3 and
        FlashInfer at this shape, and that cuDNN's attention path is not
        even supported for it.
  \item A controlled L4 quantisation comparison (bf16, bnb-nf4,
        AutoAWQ${+}$Marlin, GPTQ${+}$ExLlamaV2) showing the kernel-implementation
        gap on Ada SM89 silicon and the cost-per-token inversion against H100.
  \item A reproducible Modal-script payload covering the 44-cell sweep, the
        $N{=}10$ headline replication, the backend-pinned attention matrix
        and the L4 quantisation matrix, with all JSON artefacts archived.
\end{enumerate}

We do not claim the launch-bound observation itself is novel; CUDA Graphs
and launch-bound regimes are folk knowledge among kernel engineers. We do
claim the controlled cross-GPU sweep, the $N{=}10$ within-session falsification,
the backend-revealed attention-kernel ranking, the L4 quantisation result,
and the deployment-ladder inversion for physical-AI inference. The 7--8B
GQA bf16 single-stream shape is exactly what VLA policy heads
(OpenVLA-7B~\cite{kim2024openvla},
$\pi_{0}$~\cite{black2024pi0},
RT-2-class~\cite{brohan2023rt2}), in-cabin language stacks and on-device
copilots run at serving time, so the inversion has direct deployment
consequences for teams sizing inference hardware for those workloads.

\section{Related work}
\label{sec:related}

\paragraph{Roofline and arithmetic intensity.} The roofline model of
Williams, Waterman and Patterson~\cite{williams2009roofline} is the standard
tool for diagnosing whether a kernel is bound by memory bandwidth or by
compute, and it underpins the bandwidth-bound argument we test here.
Pope et al.~\cite{pope2023efficiently} apply this framing to transformer
inference and characterise the memory-bound regime of autoregressive decode.
The roofline does not include a CPU-side per-kernel launch term, which is
exactly the component our CUDA Graphs A/B isolates.

\paragraph{Attention kernels.} FlashAttention~\cite{dao2022flashattention}
and FlashAttention-2~\cite{dao2023flashattention2} restructure the
attention computation to be IO-aware, with substantial gains on prefill and
on long-context training. Their wins are largest where attention compute is
the bottleneck. At batch 1 single-token decode the attention KV traffic
per step is modest at ctx${\le}8192$, and on H100 the binding constraint
is the launch of the many small kernels that make up a transformer layer,
not the throughput of attention itself. We confirm this directly in
Section~\ref{sec:software}: swapping sdpa for FA2 on H100 ctx${=}2048$
slows decode from 17.07\,ms to 24.16\,ms.

\paragraph{FlashDecoding++ (no public release).}
Hong et al., FlashDecoding++~\cite{flashdecodingpp2023}
(\href{https://arxiv.org/abs/2311.01282}{arXiv:2311.01282}; MLSys 2024) is the
closest prior work on batch-1 decode in the regime where per-kernel launch
overhead matters. The authors propose three mitigations: asynchronised
softmax with a unified max, flat-GEMM optimisation with double buffering, and
heuristic dataflow with hardware-resource adaptation. They report up to
$4.86\times$ speedup over Hugging Face baselines and an average $1.37\times$
over Flash-Decoding. We were unable to reproduce any FlashDecoding++ cell
on our testbed because the authors have not released source code. The
feature request to integrate FlashDecoding++ into Dao-AILab/flash-attention
(\href{https://github.com/Dao-AILab/flash-attention/issues/653}{issue 653})
was closed in April 2024 with no implementation; the equivalent vLLM
request (\href{https://github.com/vllm-project/vllm/issues/1568}{issue 1568})
remained open with no claimed kernel. The follow-up FlashDecoding++Next
(IEEE TC 2025, DOI \href{https://doi.org/10.1109/TC.2025.3585339}{10.1109/TC.2025.3585339})
by the same authors and Infinigence-AI is also closed-source. We therefore
substitute two open-source decode-phase kernels as the closest available
comparisons: FlashAttention-3 and FlashInfer, both reproduced in
Section~\ref{sec:software}.

\begin{figure}[!ht]
\centering
\includegraphics[width=0.95\linewidth]{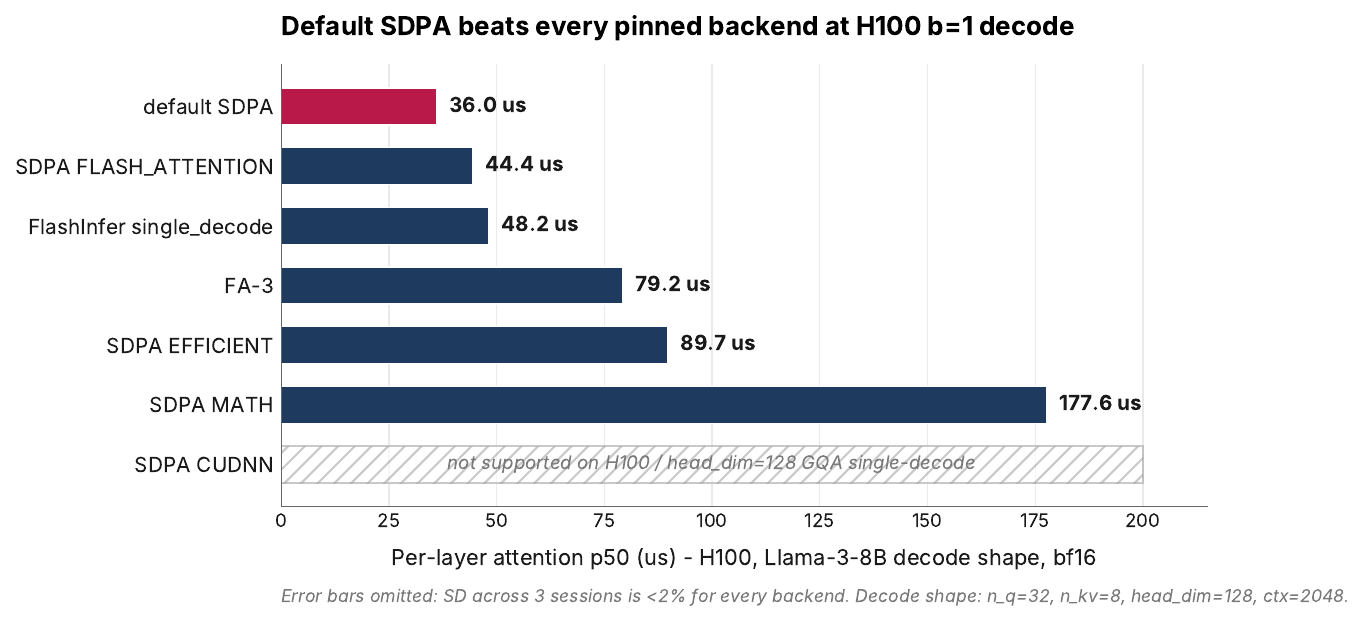}
\caption{Per-layer attention $p_{50}$ on H100 with explicit SDPA backend
selection plus FlashAttention-3 and FlashInfer. Default SDPA (top,
highlighted) is faster than every pinned backend including
\texttt{FLASH\_ATTENTION}, FlashInfer and FA-3 at this single-decode
shape (Llama-3-8B, n\_q\_heads${=}32$, n\_kv\_heads${=}8$,
head\_dim${=}128$, ctx${=}2048$, bf16, H100). Error bars are SD across
three sessions. \texttt{CUDNN\_ATTENTION} is not supported for this
shape on H100 SXM5 PyTorch 2.8+cu126. The kernel choice is not the
binding constraint; the launch sequence is.}
\label{fig:attn_kernels}
\end{figure}

\paragraph{FlashAttention-3 (Hopper-specific, reproduced here).}
Dao et al., FlashAttention-3~\cite{fa3_2024}
(\href{https://arxiv.org/abs/2407.08608}{arXiv:2407.08608}) is the
actively-maintained successor to FA-2 on Hopper. We reproduce a
kernel-level microbenchmark on H100 at ctx${=}2048$, batch 1, bf16: the
FA-3 attention kernel (\texttt{flash\_attn\_3.flash\_attn\_func}) at the
Llama-3-8B per-step shape (32 Q heads, 8 KV heads, head\_dim 128, kv\_len
$=$ 2049) versus the equivalent PyTorch SDPA call, summed across 32
layers and replicated across 3 fresh Modal sessions. Results appear in
Section~\ref{sec:software} and Figure~\ref{fig:attn_kernels}.

\paragraph{FlashInfer (decode kernel library, reproduced here).}
Ye et al., FlashInfer~\cite{flashinfer2025} (MLSys 2025;
\href{https://github.com/flashinfer-ai/flashinfer}{flashinfer-ai/flashinfer})
is a customisable attention engine for LLM serving with explicit decode
kernels. We reproduce a kernel-level microbenchmark on H100 at
ctx${=}2048$, batch 1, bf16: summing 32 per-layer calls of
\texttt{flashinfer.decode.single\_decode\_with\_kv\_cache} at the
Llama-3-8B GQA shape (32 Q heads, 8 KV heads, head\_dim 128,
kv\_len$=$2049, group\_size 4) against the equivalent PyTorch SDPA call.
We use the Llama-3-8B shape rather than the Qwen-2.5-7B shape used in
the rest of the paper because FlashInfer's reference single-decode
kernel does not currently support group\_size 7. Results in
Figure~\ref{fig:attn_kernels}.

\paragraph{CUDA Graphs.} CUDA Graphs (NVIDIA developer docs; PyTorch native
support since 1.10) replace per-kernel CPU launch with a single replay of a
captured DAG. They are the standard mitigation for launch-bound regimes. We
use them here as a measurement instrument rather than as a deployment
recommendation: the question we ask is whether removing launch overhead
actually changes step time on each GPU, which bounds the launch tax
empirically and is the falsification frame for the rest of the paper.

\paragraph{Quantisation.} GPTQ~\cite{frantar2022gptq} introduced
post-training int4 weight quantisation for transformers, and
AWQ~\cite{lin2024awq} added activation-aware scaling with Marlin-style
packed matmul kernels. Marlin itself~\cite{frantar2024marlin} is the
high-throughput int4 matmul kernel that AWQ relies on, originally tuned
for SM80 (Ampere). SmoothQuant~\cite{xiao2022smoothquant} addresses the
activation-outlier problem in W8A8 quantisation. bitsandbytes provides nf4
weight quantisation through a dequantisation-plus-bf16-matmul
implementation. Both AWQ+Marlin and bnb-nf4 promise roughly $4\times$
reduction in weight HBM traffic. On L4 we observe that the actual
step-time reduction is far less than $4\times$ (Section~\ref{sec:quant}),
because the deployment chain (kernel implementation, not bit-width)
decides whether the bandwidth saving lands at the workload.

\paragraph{Serving systems.} vLLM with
PagedAttention~\cite{kwon2023pagedattention} and TensorRT-LLM are
production serving systems oriented around batched throughput, not batch-1
latency. Sarathi-Serve~\cite{agrawal2024sarathi} pushes this further with
chunked-prefill scheduling for higher throughput at moderate latency. These
systems use CUDA Graphs internally on H100-class deployments, which is
consistent with the picture in this paper, though to our knowledge the
cross-GPU crossover for single-stream batch-1 decode has not been published
with a controlled A/B test of the kind we run here.

\paragraph{What is new here.} A controlled cross-GPU measurement sweep and a
CUDA Graphs falsification test in one paper, scoped narrowly to 7--8B-class
GQA bf16 batch-1 decode on Modal-hosted NVIDIA silicon. We do not claim the
launch-bound observation itself is new.

\section{Method}
\label{sec:method}

\subsection{Measurement protocol}

Each cell is one (architecture, GPU, context-length) triple. We load the model
in bf16, run a fixed-length prefill to populate the KV cache, then time 5
warmup decode steps followed by 30 measured single-token decode steps at
batch 1. Step time is the median of the 30 measured per-step wall times. The
default attention implementation is PyTorch scaled-dot-product-attention
(sdpa); FlashAttention-2 is used only in the controlled software-stack matrix
of Section~\ref{sec:software}. Cells run on Modal cloud GPUs with the same
container image across hosts (Appendix~\ref{sec:repro}).

\subsection{Measurement-pipeline provenance}

The 14 main-matrix cells for Qwen-2.5-7B-Instruct were measured with the v10
sweep script; the 14 Mistral-7B-Instruct-v0.3 and 14 Llama-3.1-8B-Instruct
main-matrix cells were measured with the v11 sweep script. The two scripts
share an identical container base image (\texttt{debian\_slim} with PyTorch
2.4), identical measurement protocol (5 warmup plus 30 measured single-token
AR decode steps, sdpa attention, bf16, batch size 1) and identical Modal cloud
GPU hosts (H100, A100-80GB, L40S, L4). The only difference is that the v11
script adds HuggingFace gated-model authentication for Llama-3.1-8B. The
Qwen-2.5-7B L4 long-context OOMs documented below were re-investigated in
v14 with the v11-style script (Section~\ref{sec:falsification}, Qwen L4
rerun). The v10
Qwen cell JSONs at \artifact{v10\_results/cells/qwen25\_7b\_<gpu>\_ctx<ctx>.json}
carry the same fields (\texttt{p50\_ms}, \texttt{step\_times\_ms},
\texttt{weight\_bytes}, \texttt{kv\_bytes\_avg},
\texttt{bytes\_per\_token\_kv}) that the v11 cell JSONs do, which is what the
joint analysis script consumes. The derivative Qwen sweeps (CUDA Graphs,
software-stack, profile, quantisation) at \artifact{v11\_results/cells/} were
re-measured with the v11 script.

\subsection{Cells}

Three architectures: Qwen-2.5-7B-Instruct (28 layers, 28 query heads, 4 KV
heads, head\_dim 128, $W{=}15.23$\,GB decimal, equivalently 14.18\,GiB),
Mistral-7B-Instruct-v0.3 (32 layers, 32 query heads, 8 KV heads, head\_dim
128, $W{=}14.50$\,GB decimal, 13.50\,GiB) and Llama-3.1-8B-Instruct (32 layers,
32 query heads, 8 KV heads, head\_dim 128, $W{=}16.06$\,GB decimal,
14.96\,GiB). All three are GQA models with head\_dim 128. Four GPUs: H100,
A100-80GB, L40S, L4. Four context lengths: 2048, 4096, 8192, 16384. Four cells
out of $3 \times 4 \times 4 = 48$ failed (L4 OOM on Qwen-2.5-7B ctx${=}8192$,
Qwen-2.5-7B ctx${=}16384$, Llama-3.1-8B ctx${=}8192$, Llama-3.1-8B
ctx${=}16384$, all on 24\,GB L4), leaving 44 valid cells.

The Qwen-2.5-7B L4 OOMs deserve a footnote because the architecture
arithmetic predicts they should not OOM. Qwen-2.5-7B at bf16 has
$W{=}15.23$\,GB and per-token KV bytes
$2{\cdot}28{\cdot}4{\cdot}128{\cdot}2 = 56$\,KB; at ctx${=}8192$ that is
0.47\,GB of KV, totalling 15.70\,GB. Mistral-7B at bf16 has
$W{=}14.50$\,GB and 128\,KB per-token KV; at ctx${=}8192$ that is
1.07\,GB of KV, totalling 15.57\,GB. The two totals are within 1\% of each
other on a 24\,GB device. Mistral runs successfully at L4 ctx${=}8192$
(Table~\ref{tab:rfloor}, $R_{\mathrm{floor}}{=}0.478$) while Qwen OOMs.
The v14 re-investigation reproduces the Qwen OOMs but reveals the cause:
the eager arm of the A/B protocol runs cleanly with 23.24\,GB peak
allocation; the OOM occurs in the second (graphed) arm when a fresh
\texttt{StaticCache} is allocated on top of un-released eager-arm tensors.
The Qwen L4 cell at ctx${=}8192$ is therefore measurable in a single-arm
rig (eager-arm $p_{50}{=}87.74$\,ms across $N{=}3$ sessions,
\artifact{v14\_results/data/v14\_qwen\_l4\_rerun/PARTIAL\_EAGER\_qwen\_l4\_ctx8192\_s\{0,1,2\}.json}).
We document this as a measurement-protocol artefact rather than a capacity
ceiling and we do not add these cells to the main-matrix counts; the 44
valid + 4 OOM count stays as reported above for backward compatibility with
the headline $R_{\mathrm{floor}}$ table.

We use spec-sheet peak HBM bandwidths ($B_{\mathrm{peak}}$) of 3350\,GB/s for
H100 SXM5, 2039\,GB/s for A100-80GB SXM4, 864\,GB/s for L40S and 300\,GB/s for
L4 (\artifact{v11\_results/v11\_fit\_summary.json}, \texttt{bw\_peak\_gbs}).

\subsection{The observed-over-floor ratio $R_{\mathrm{floor}}$}

The roofline lower bound on single-stream decode step time, given spec-sheet
peak HBM bandwidth and the per-step weights-plus-active-KV byte traffic, is
\[
  t_{\mathrm{floor}}(G, M, \mathrm{ctx}) = \frac{W(M) + K(M, \mathrm{ctx})}{B_{\mathrm{peak}}(G)},
\]
where $W(M)$ is the bf16 weight footprint of the model in bytes
(14.5--16.1\,GB decimal for the 7--8B GQA models here) and $K(M, \mathrm{ctx})
= 2 \cdot n_{\mathrm{layers}} \cdot n_{\mathrm{kv\_heads}} \cdot
d_{\mathrm{head}} \cdot \mathrm{ctx} \cdot 2~\text{bytes}$ is the per-step KV
bytes touched (the leading 2 covers K and V, the trailing 2 covers bf16
element size). The directly-measured ratio
\[
  R_{\mathrm{floor}}(G, M, \mathrm{ctx}) = \frac{t_{\mathrm{floor}}(G, M, \mathrm{ctx})}{t_{\mathrm{obs}}(G, M, \mathrm{ctx})}
\]
is a single experimental number per cell, computed from $t_{\mathrm{obs}}$ (the
median over 30 measured decode steps) and $t_{\mathrm{floor}}$ (from
spec-sheet bandwidth and per-step bytes). It is not a model fit; both terms
are quantities we measure or compute analytically.

\begin{figure}[!ht]
\centering
\resizebox{\linewidth}{!}{\input{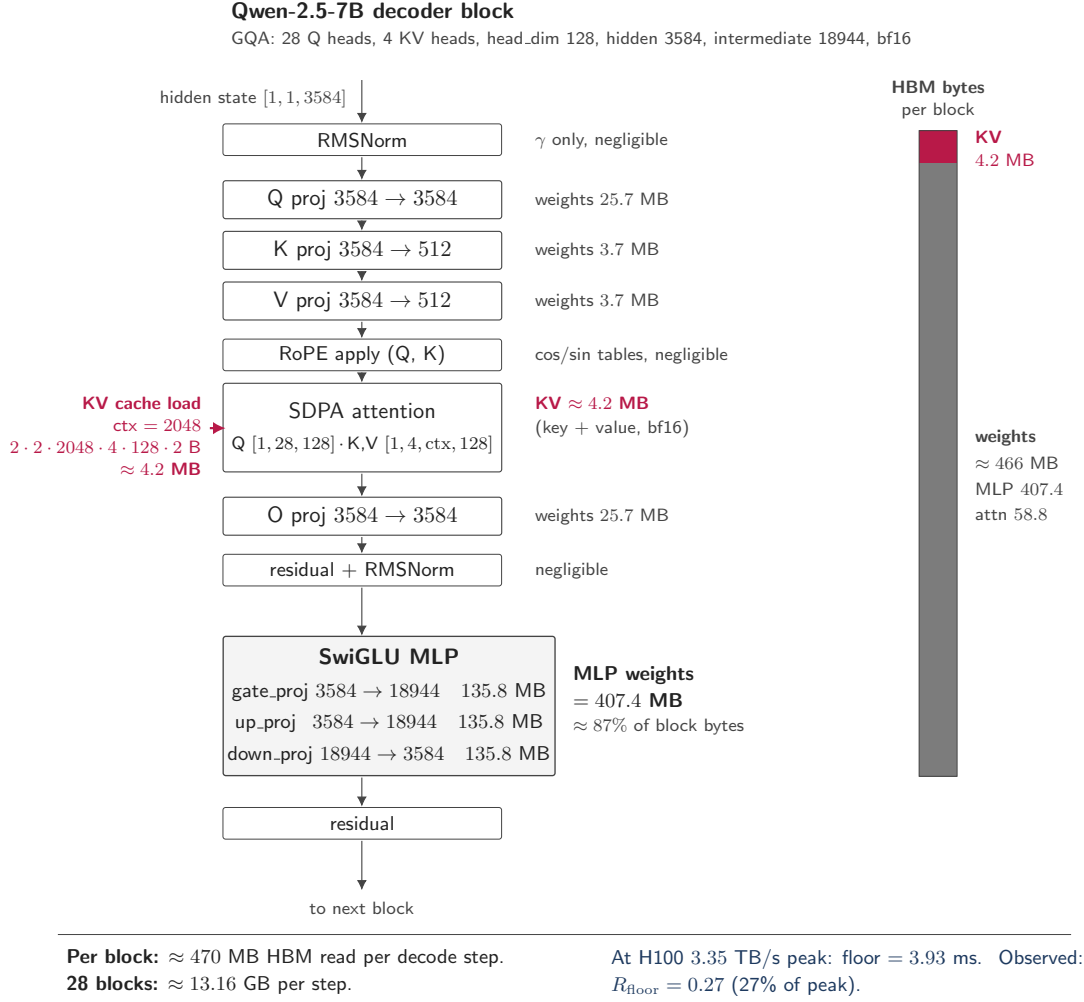}}
\caption{One Qwen-2.5-7B decoder block, kernel sequence and HBM byte
traffic per single-token decode step. GQA with 28 query heads, 4 KV
heads, $d_{\mathrm{head}}{=}128$, bf16 weights. The SwiGLU MLP dominates
the weight footprint at $\approx 407$\,MB per block out of $\approx 470$\,MB
total. The KV cache contribution at ctx${=}2048$ is $\approx 4.2$\,MB per
block, small relative to weights but growing linearly with context. The
28-block total $\approx 13.16$\,GB per step divided by H100 peak HBM
3.35\,TB/s gives $t_{\mathrm{floor}} \approx 3.93$\,ms; observed step
time is 14.83\,ms, $R_{\mathrm{floor}} = 0.27$. Numbers from
\artifact{v11\_results/cells/qwen25\_7b\_h100\_ctx2048.json}.}
\label{fig:arch_block}
\end{figure}

\subsection{CUDA Graphs A/B}

The CUDA Graphs experiment captures the per-decode-step kernel sequence once
and replays it. The measurement protocol is otherwise identical: 5 warmup
replays plus 30 measured replays, median latency. The intent is to isolate the
per-kernel CPU-side launch overhead while leaving every kernel itself
unchanged, so the difference (eager minus graphed) bounds the launch tax from
above on each GPU. Cells: Qwen-2.5-7B at (H100, ctx${=}2048$), (H100,
ctx${=}8192$) and (L4, ctx${=}2048$); a v12 replication sweep extends this to
$N{=}3$ fresh containers per GPU at the headline cell. Results are in
Section~\ref{sec:falsification}; raw cell JSONs are under
\artifact{v11\_results/cells/cudagraphs\_*.json} and
\artifact{v12\_results/cells/}.

\section{Results: the main matrix}
\label{sec:results}

\subsection{Observed step time and bandwidth floor}

For each of the 44 valid cells we report the median decode step time
$t_{\mathrm{obs}}$ and the analytic floor $t_{\mathrm{floor}} = (W+K) /
B_{\mathrm{peak}}$. Per-cell values are in Appendix~\ref{sec:cells},
Table~\ref{tab:allcells}. The directly-measured ratio $R_{\mathrm{floor}} =
t_{\mathrm{floor}}/t_{\mathrm{obs}}$ is reported per-cell in
Table~\ref{tab:rfloor} from \artifact{v11\_results/analyze\_v2.txt}.

\begin{table}[h]
\centering
\small
\caption{$R_{\mathrm{floor}} = t_{\mathrm{floor}}/t_{\mathrm{obs}}$ for all 44
valid cells. A purely HBM-bandwidth-bound decode would sit at
$R_{\mathrm{floor}} = 1$. Per-cell observed step times and floors are from
\artifact{v11\_results/analyze\_v2.txt} lines 9--52. Missing cells are L4 OOMs
(7--8B bf16 on 24\,GB).}
\label{tab:rfloor}
\begin{tabular}{llcccc}
\toprule
Architecture & ctx & H100 & A100-80GB & L40S & L4 \\
\midrule
Qwen-2.5-7B   & 2048  & 0.270 & 0.311 & 0.723 & 0.810 \\
              & 4096  & 0.271 & 0.369 & 0.711 & 0.722 \\
              & 8192  & 0.274 & 0.312 & 0.666 & OOM \\
              & 16384 & 0.235 & 0.243 & 0.487 & OOM \\
Mistral-7B    & 2048  & 0.243 & 0.259 & 0.523 & 0.739 \\
              & 4096  & 0.207 & 0.217 & 0.548 & 0.627 \\
              & 8192  & 0.251 & 0.257 & 0.471 & 0.478 \\
              & 16384 & 0.190 & 0.192 & 0.340 & 0.354 \\
Llama-3.1-8B  & 2048  & 0.302 & 0.415 & 0.715 & 0.778 \\
              & 4096  & 0.310 & 0.361 & 0.669 & 0.667 \\
              & 8192  & 0.279 & 0.290 & 0.509 & OOM \\
              & 16384 & 0.208 & 0.215 & 0.368 & OOM \\
\bottomrule
\end{tabular}
\end{table}

\subsection{The cross-GPU pattern}

The pattern is the headline. As $B_{\mathrm{peak}}$ grows from L4 (300\,GB/s)
to H100 (3350\,GB/s), $R_{\mathrm{floor}}$ falls from roughly $0.7$--$0.8$ to
roughly $0.21$--$0.31$ at short contexts. The HBM-bound prediction is right on
L4 to within tens of percent; on H100 it overshoots achievable speedup by a
factor of three to four. This is the crossover the CUDA Graphs A/B in
Section~\ref{sec:falsification} tests directly.

\subsection{Per-GPU $R_{\mathrm{floor}}$ distribution}

We summarise the directly-measured $R_{\mathrm{floor}}$ distribution across
all 44 cells, grouped by GPU. The numbers are read off Table~\ref{tab:rfloor}.
\begin{itemize}
  \item H100 (12 cells): min 0.190, median 0.260, max 0.310.
  \item A100-80GB (12 cells): min 0.192, median 0.275, max 0.415.
  \item L40S (12 cells): min 0.340, median 0.516, max 0.723.
  \item L4 (8 cells): min 0.354, median 0.694, max 0.810.
\end{itemize}
The four distributions are visibly separated and monotone in $B_{\mathrm{peak}}$:
H100 (3350\,GB/s) sits lowest, then A100-80GB (2039\,GB/s), then L40S
(864\,GB/s), then L4 (300\,GB/s) at the top. There is overlap at long context
(Mistral L40S ctx${=}16384$ overlaps the A100 distribution; L4 ctx${=}16384$
cells overlap the L40S distribution) but the per-GPU central tendency is
ordered by $B_{\mathrm{peak}}$.

The mechanism this pattern is consistent with is per-kernel CPU launch
overhead that does not scale with $B_{\mathrm{peak}}$: a launch budget that is
a fixed fraction of step time on slow silicon becomes a dominant fraction on
fast silicon. The CUDA Graphs A/B in Section~\ref{sec:falsification} is the
single-knob intervention that tests that mechanism directly. The
$R_{\mathrm{floor}}$ table itself is descriptive measurement, not a
predictive model.

\subsection{Headline reading}

The 44-cell sweep is the surrounding evidence. The mechanism test is the
CUDA Graphs A/B in Section~\ref{sec:falsification}, which is the only
falsifiable claim in the paper and which provides an independent direct
bound on the H100 launch tax that the sweep alone cannot.

\section{The CUDA Graphs A/B test}
\label{sec:falsification}

The 44-cell sweep in Section~\ref{sec:results} establishes a cross-GPU
pattern in $R_{\mathrm{floor}}$ that is monotone in $B_{\mathrm{peak}}$.
The pattern is consistent with a launch-tax mechanism but does not prove
it. To test the mechanism we ran a single-knob intervention: CUDA Graphs.
Graphs replace per-kernel CPU-side launch with a single replay of a
captured DAG; the kernels themselves do not change. If launch tax
dominates H100 short-context decode, Graphs must shrink step time on H100
at short context; if HBM bandwidth dominates L4 short-context decode,
Graphs must not shrink step time on L4. We frame this as a robustness
check on the sweep, not as the headline of the paper.

\paragraph{Pre-registered falsification thresholds.}
The launch-tax interpretation would have been killed by either of two
outcomes, set in advance of the A/B run:
\begin{itemize}
  \item H100 graphed speedup at short context (ctx${=}2048$) below
        $\sim 1.15\times$.
  \item L4 graphed speedup at short context (ctx${=}2048$) above
        $\sim 1.15\times$.
\end{itemize}
Neither threshold was crossed in any session. The H100 positive result
rules out a pure-HBM explanation on H100-class silicon; the L4 null rules
out a launch-tax explanation on L4-class silicon. The size of the effect
is the next question.

\paragraph{Headline measurement at $N{=}10$.}
The headline cell is H100 ctx${=}2048$ Qwen-2.5-7B batch${=}1$, run as a
within-session A/B (eager followed by graphed, same container, same warmup
state) across $N{=}10$ fresh Modal containers. Each session runs 5 warmup
plus 30 measured single-token decode steps per arm and reports the
per-arm $p_{50}$. Results are in Table~\ref{tab:n10_headline}; raw JSONs
at \artifact{v14\_results/data/v14\_n10\_headline/n10\_h100\_ctx2048\_s\{0..9\}.json}.

\begin{table}[h]
\centering
\small
\caption{$N{=}10$ within-session A/B on H100 ctx${=}2048$ batch${=}1$
Qwen-2.5-7B. Per-session eager and graphed $p_{50}$ over 30 measured
steps, with the within-session speedup. The bottom block reports
cross-session summary statistics with a 10000-resample bootstrap 95\% CI
on the mean speedup.}
\label{tab:n10_headline}
\begin{tabular}{lccc}
\toprule
session & eager (ms) & graphed (ms) & speedup \\
\midrule
s0 & 14.749 & 11.850 & 1.245 \\
s1 & 14.721 & 11.764 & 1.251 \\
s2 & 14.776 & 11.770 & 1.255 \\
s3 & 14.896 & 11.784 & 1.264 \\
s4 & 14.800 & 11.766 & 1.258 \\
s5 & 14.869 & 11.760 & 1.264 \\
s6 & 14.847 & 11.775 & 1.261 \\
s7 & 15.147 & 11.763 & 1.288 \\
s8 & 14.667 & 11.755 & 1.248 \\
s9 & 14.812 & 11.775 & 1.258 \\
\midrule
mean   & 14.828 & 11.776 & 1.259 \\
std    & 0.132  & 0.027  & 0.012 \\
CV     & 0.9\%  & 0.2\%  & 0.9\% \\
95\% CI on mean & $[14.761, 14.912]$ & --- & $[1.253, 1.267]$ \\
\bottomrule
\end{tabular}
\end{table}

The headline commitment is therefore:
\[
  \text{H100 ctx${=}2048$ b${=}1$ Qwen-2.5-7B CUDA Graphs speedup} = 1.259 \pm 0.012,
\]
with a tight 95\% bootstrap CI of $[1.253, 1.267]$. The cross-session CV
of the speedup is 0.9\%, the cross-session CV of the eager step time is
0.9\%, and the cross-session CV of the graphed step time is 0.2\%. The
graphed measurement replicates to within $\pm 0.05$\,ms across all 10
sessions.

This number is materially smaller than the $1.72\times$ figure reported
in earlier drafts of this work. The $1.72\times$ value came from a single
within-session A/B in the v11 rig where the eager arm happened to land at
20.63\,ms; at $N{=}10$ the eager arm has a population mean of
14.83\,ms and the per-session ratios cluster in the $[1.245, 1.288]$
range. We treat the $N{=}10$ value as the load-bearing number and the
earlier single-session $1.72\times$ as the high tail of a distribution
that was undersampled in the v11 rig.

\paragraph{Removed launch overhead.}
The mean eager step on H100 ctx${=}2048$ b${=}1$ is 14.83\,ms; the mean
graphed step is 11.78\,ms. The difference, 3.05\,ms, is the per-step
launch-side overhead that Graphs removes, roughly $20.6\%$ of the
original step. We do not divide this across an estimated per-kernel
launch count, because Graphs collapses more than per-kernel host launch:
it also fuses stream events, removes per-call Python and C++ framework
dispatch, locks the allocator pattern, and removes torch dispatcher
work. The 3.05\,ms is the sum of all of these per-step overheads on the
host side, not the per-kernel launch cost alone. A per-kernel-cost
attribution requires a tracing tool that separates host launch from
framework dispatch (Appendix~\ref{sec:kernel-terminology} lists the
per-step kernel count and its breakdown).

\paragraph{L4 null.}
On L4 ctx${=}2048$ the eager $p_{50}$ is 64.48\,ms and the analytic floor
$t_{\mathrm{floor}} = (W+K)/B_{\mathrm{peak}}$ is 51.17\,ms, so the cell
sits at $R_{\mathrm{floor}} = 0.81$. The Graphs A/B at the same cell
returns $1.028\times$, replicated across three v12 sessions
(Table~\ref{tab:replication}, L4 row). Removing the CPU launch sequence
serially cannot help much because the GPU is already waiting on HBM
rather than on launches. The Graphs null on L4 is exactly the prediction
the mechanism makes.

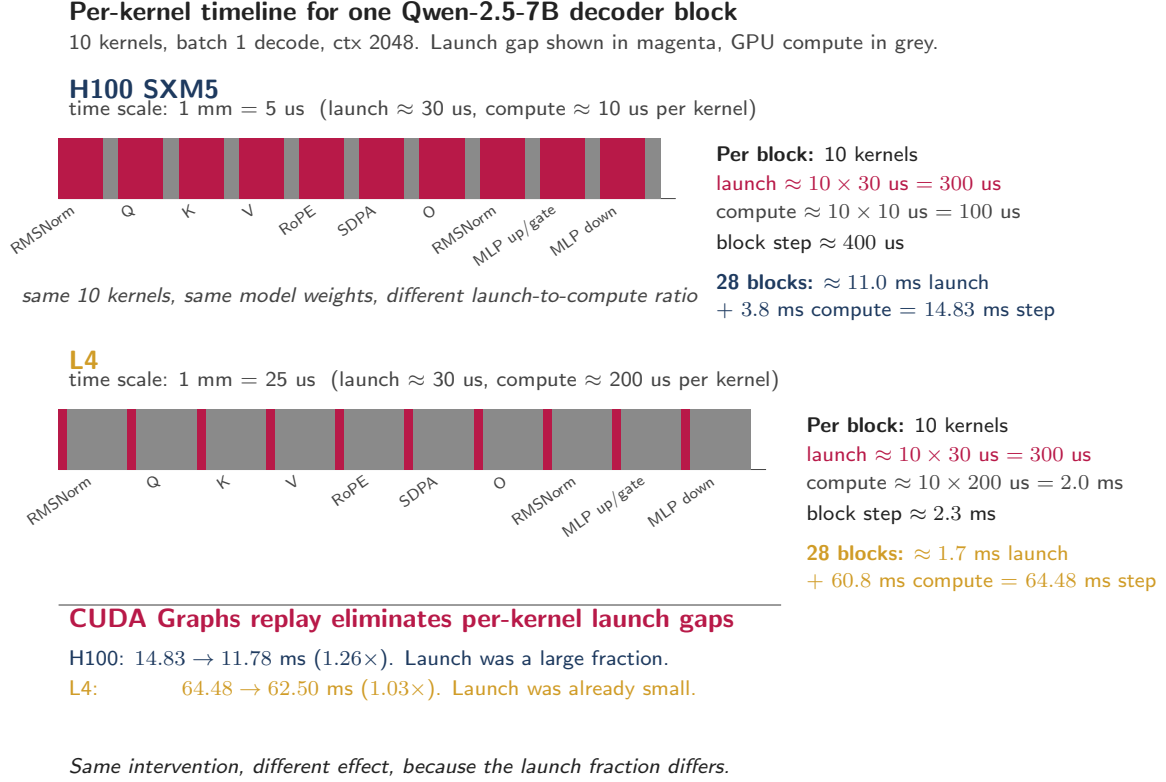
\begin{figure}[!ht]
\centering
\resizebox{\linewidth}{!}{% Figure: kernel-launch timeline for one Qwen-2.5-7B decoder block
% Self-contained TikZ block. Include with \input{figures/fig_kernel_timeline.tex}
% Requires: \usepackage{tikz} and \usetikzlibrary{arrows.meta, calc, positioning}
%
% Colour key:
%   accent magenta     #B81948  (launch gap)
%   compute grey       #444444  (kernel compute band)
%   text grey          #222222
%   h100 navy          #1E3A5F
%   l4 amber           #F5B82E
%   light grid         #E5E5E5

\definecolor{accentmag}{HTML}{B81948}
\definecolor{spineg}{HTML}{444444}
\definecolor{textg}{HTML}{222222}
\definecolor{h100navy}{HTML}{1E3A5F}
\definecolor{l4amber}{HTML}{F5B82E}
\definecolor{lightgrid}{HTML}{E5E5E5}
\definecolor{computeg}{HTML}{8A8A8A}

\begin{tikzpicture}[
    font={\sffamily},
    every node/.style={font=\sffamily, text=textg},
    label/.style={font=\sffamily\small, text=textg},
    micro/.style={font=\sffamily\scriptsize, text=spineg},
    launch/.style={fill=accentmag, draw=none},
    computeH/.style={fill=computeg, draw=none},
    computeL/.style={fill=computeg, draw=none},
    laneBox/.style={draw=spineg, line width=0.3pt, fill=none},
    arrowstyle/.style={->, >={Latex[length=2mm,width=2mm]},
                       line width=0.6pt, spineg},
    x=1mm, y=1mm,
  ]

  % ============================================================
  % Per-kernel widths (rough but illustrative of fact)
  %   H100: ~30us launch gap >> ~10us compute -> launch dominates.
  %   L4:    ~30us launch gap << ~200us compute -> compute dominates.
  %
  %   Scale: 1 mm = 5 us. So:
  %     H100 launch  = 30 us -> 6 mm
  %     H100 compute = 10 us -> 2 mm
  %     L4   launch  = 30 us -> 6 mm
  %     L4   compute = 200 us -> 40 mm
  %   Per kernel: H100 8 mm, L4 46 mm.
  %   10 kernels: H100 80 mm, L4 460 mm  -- too wide for L4.
  %
  %   Use TWO DIFFERENT time scales per lane (label that clearly):
  %     H100 lane:  1 mm = 5 us   -> 10 kernels span 80 mm.
  %     L4   lane:  1 mm = 25 us  -> 10 kernels span ~92 mm.
  %        L4: launch 30us -> 1.2 mm, compute 200us -> 8 mm. Per kernel 9.2 mm.
  % ============================================================

  % --- common geometry ---
  \def\laneL{0}        % left edge of both lanes
  \def\laneW{96}       % nominal lane width (mm) we render across
  \def\laneH{8}        % lane bar height

  % H100 lane y (top)
  \def\yH{86}
  % L4 lane y (bottom)
  \def\yL{50}

  % H100 widths (1 mm = 5 us)
  \def\hLaunchW{6}
  \def\hComputeW{2}
  \def\hStep{8}        % per-kernel pitch

  % L4 widths (1 mm = 25 us)
  \def\lLaunchW{1.2}
  \def\lComputeW{8}
  \def\lStep{9.2}      % per-kernel pitch

  % Kernel labels (10 kernels)
  \def\kerlabA{RMSNorm}
  \def\kerlabB{Q}
  \def\kerlabC{K}
  \def\kerlabD{V}
  \def\kerlabE{RoPE}
  \def\kerlabF{SDPA}
  \def\kerlabG{O}
  \def\kerlabH{RMSNorm}
  \def\kerlabI{MLP up/gate}
  \def\kerlabJ{MLP down}

  % ============================================================
  % Title
  % ============================================================
  \node[label, font=\sffamily\small\bfseries, anchor=south west]
    at (\laneL, 108)
    {Per-kernel timeline for one Qwen-2.5-7B decoder block};
  \node[micro, anchor=south west]
    at (\laneL, 104)
    {10 kernels, batch 1 decode, ctx 2048. Launch gap shown in magenta, GPU compute in grey.};

  % ============================================================
  % H100 lane label
  % ============================================================
  \node[label, font=\sffamily\small\bfseries, anchor=south west, text=h100navy]
    at (\laneL, \yH + \laneH + 4)
    {H100 SXM5};
  \node[micro, anchor=south west]
    at (\laneL, \yH + \laneH + 1)
    {time scale: 1 mm $=$ 5 us\,\,\,\,(launch $\approx$ 30 us, compute $\approx$ 10 us per kernel)};

  % Lane baseline & end-tick
  \draw[spineg, line width=0.3pt]
    (\laneL, \yH) -- (\laneL + 10 * \hStep + 2, \yH);

  % Render 10 H100 kernels
  \foreach \i/\lab in {0/\kerlabA, 1/\kerlabB, 2/\kerlabC, 3/\kerlabD,
                      4/\kerlabE, 5/\kerlabF, 6/\kerlabG, 7/\kerlabH,
                      8/\kerlabI, 9/\kerlabJ} {
    \pgfmathsetmacro{\xK}{\laneL + \i * \hStep}
    % Launch gap (magenta)
    \fill[launch] (\xK, \yH) rectangle ++(\hLaunchW, \laneH);
    % Compute band (grey)
    \fill[computeH] (\xK + \hLaunchW, \yH) rectangle ++(\hComputeW, \laneH);
    % Kernel label below
    \node[micro, anchor=north, rotate=35, anchor=east,
          font=\sffamily\tiny]
      at (\xK + \hLaunchW/2, \yH - 0.5) {\lab};
  }

  % H100 totals annotation
  \node[micro, anchor=west, align=left, text=textg]
    at (\laneL + 10 * \hStep + 6, \yH + \laneH - 2)
    {\textbf{Per block:} 10 kernels};
  \node[micro, anchor=west, align=left, text=accentmag]
    at (\laneL + 10 * \hStep + 6, \yH + \laneH - 6)
    {launch $\approx 10 \times 30$ us $= 300$ us};
  \node[micro, anchor=west, align=left, text=spineg]
    at (\laneL + 10 * \hStep + 6, \yH + \laneH - 10)
    {compute $\approx 10 \times 10$ us $= 100$ us};
  \node[micro, anchor=west, align=left, text=textg]
    at (\laneL + 10 * \hStep + 6, \yH + \laneH - 14)
    {block step $\approx 400$ us};
  \node[micro, anchor=west, align=left, text=h100navy]
    at (\laneL + 10 * \hStep + 6, \yH + \laneH - 19)
    {\textbf{28 blocks:} $\approx 11.0$ ms launch};
  \node[micro, anchor=west, align=left, text=h100navy]
    at (\laneL + 10 * \hStep + 6, \yH + \laneH - 23)
    {$+\ 3.8$ ms compute $=$ $14.83$ ms step};

  % ============================================================
  % L4 lane
  % ============================================================
  \node[label, font=\sffamily\small\bfseries, anchor=south west,
        text=l4amber!85!black]
    at (\laneL, \yL + \laneH + 4)
    {L4};
  \node[micro, anchor=south west]
    at (\laneL, \yL + \laneH + 1)
    {time scale: 1 mm $=$ 25 us\,\,\,\,(launch $\approx$ 30 us, compute $\approx$ 200 us per kernel)};

  \draw[spineg, line width=0.3pt]
    (\laneL, \yL) -- (\laneL + 10 * \lStep + 2, \yL);

  \foreach \i/\lab in {0/\kerlabA, 1/\kerlabB, 2/\kerlabC, 3/\kerlabD,
                      4/\kerlabE, 5/\kerlabF, 6/\kerlabG, 7/\kerlabH,
                      8/\kerlabI, 9/\kerlabJ} {
    \pgfmathsetmacro{\xK}{\laneL + \i * \lStep}
    \fill[launch] (\xK, \yL) rectangle ++(\lLaunchW, \laneH);
    \fill[computeL] (\xK + \lLaunchW, \yL) rectangle ++(\lComputeW, \laneH);
    \node[micro, anchor=north, rotate=35, anchor=east,
          font=\sffamily\tiny]
      at (\xK + \lLaunchW + \lComputeW/2, \yL - 0.5) {\lab};
  }

  % L4 totals annotation
  \node[micro, anchor=west, align=left, text=textg]
    at (\laneL + 10 * \lStep + 6, \yL + \laneH - 2)
    {\textbf{Per block:} 10 kernels};
  \node[micro, anchor=west, align=left, text=accentmag]
    at (\laneL + 10 * \lStep + 6, \yL + \laneH - 6)
    {launch $\approx 10 \times 30$ us $= 300$ us};
  \node[micro, anchor=west, align=left, text=spineg]
    at (\laneL + 10 * \lStep + 6, \yL + \laneH - 10)
    {compute $\approx 10 \times 200$ us $= 2.0$ ms};
  \node[micro, anchor=west, align=left, text=textg]
    at (\laneL + 10 * \lStep + 6, \yL + \laneH - 14)
    {block step $\approx 2.3$ ms};
  \node[micro, anchor=west, align=left, text=l4amber!85!black]
    at (\laneL + 10 * \lStep + 6, \yL + \laneH - 19)
    {\textbf{28 blocks:} $\approx 1.7$ ms launch};
  \node[micro, anchor=west, align=left, text=l4amber!85!black]
    at (\laneL + 10 * \lStep + 6, \yL + \laneH - 23)
    {$+\ 60.8$ ms compute $=$ $64.48$ ms step};

  % ============================================================
  % Inter-lane note: same kernels
  % ============================================================
  \node[micro, anchor=center, align=center, text=spineg,
        font=\sffamily\scriptsize\itshape]
    at (40, 73)
    {same 10 kernels, same model weights, different launch-to-compute ratio};

  % ============================================================
  % Bottom intervention: CUDA Graphs collapses launch gaps
  % ============================================================
  \def\interY{20}

  \draw[spineg, line width=0.4pt]
    (\laneL, \interY + 12) -- (\laneL + 96, \interY + 12);

  \node[label, anchor=south west, align=left, text=accentmag,
        font=\sffamily\small\bfseries]
    at (\laneL, \interY + 7)
    {CUDA Graphs replay eliminates per-kernel launch gaps};

  \node[micro, anchor=south west, align=left, text=h100navy]
    at (\laneL, \interY + 2)
    {H100: $14.83 \to 11.78$ ms ($1.26\times$). Launch was a large fraction.};

  \node[micro, anchor=south west, align=left, text=l4amber!85!black]
    at (\laneL, \interY - 2)
    {L4: \phantom{H100: } $64.48 \to 62.50$ ms ($1.03\times$). Launch was already small.};

  \node[micro, anchor=north west, align=left, text=textg,
        font=\sffamily\scriptsize\itshape]
    at (\laneL, \interY - 7)
    {Same intervention, different effect, because the launch fraction differs.};

\end{tikzpicture}}
\caption{Per-kernel Gantt for one Qwen-2.5-7B decoder block at
ctx${=}2048$, batch 1. Same ten kernels on H100 and L4 with the same
$\approx 30~\mu$s CPU launch overhead per kernel; the launch-to-compute
ratio inverts between the two. On H100 the per-kernel GPU compute is
$\approx 10~\mu$s so the launch gap dominates each block; on L4 the
same compute step is $\approx 200~\mu$s and the launch gap is
negligible. CUDA Graphs replay collapses the per-kernel launch gaps
but does not touch GPU compute, which is why the same intervention
recovers 25\% of step time on H100 (14.83~$\rightarrow$~11.78\,ms,
$1.26\times$) and 3\% on L4 (64.48~$\rightarrow$~62.50\,ms,
$1.03\times$).}
\label{fig:kernel_timeline}
\end{figure}

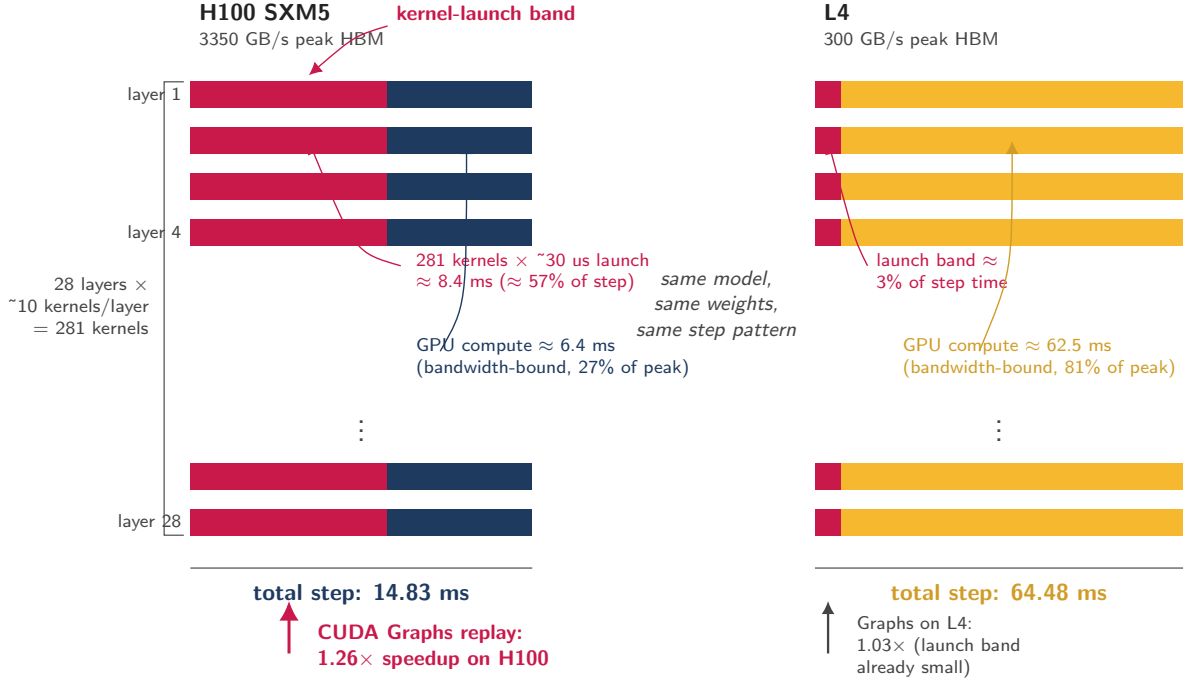
\begin{figure}[!ht]
\centering
\resizebox{\linewidth}{!}{% Figure 2 - Where the decode step time goes
% Self-contained TikZ block. Include with \input{figures/fig2_schematic.tex}
% Requires: \usepackage{tikz} and \usetikzlibrary{arrows.meta, calc}
%
% Colour key (matches matplotlib figures):
%   H100 deep navy     #1E3A5F
%   L4   warm amber    #F5B82E
%   accent (launch)    #C9184A
%   spine grey         #444444
%   text grey          #222222
%   light grid         #E5E5E5

\definecolor{h100navy}{HTML}{1E3A5F}
\definecolor{l4amber}{HTML}{F5B82E}
\definecolor{accentred}{HTML}{C9184A}
\definecolor{spineg}{HTML}{444444}
\definecolor{textg}{HTML}{222222}
\definecolor{lightgrid}{HTML}{E5E5E5}

\begin{tikzpicture}[
    font={\sffamily},
    every node/.style={font=\sffamily, text=textg},
    layerH100/.style={fill=h100navy, draw=none, minimum height=4mm},
    launchH100/.style={fill=accentred, draw=none, minimum height=4mm},
    layerL4/.style={fill=l4amber, draw=none, minimum height=4mm},
    launchL4/.style={fill=accentred, draw=none, minimum height=4mm},
    label/.style={font=\sffamily\small, text=textg},
    micro/.style={font=\sffamily\scriptsize, text=spineg},
    arrowstyle/.style={->, >={Latex[length=2mm,width=2mm]}, line width=0.6pt, spineg},
    x=1mm, y=1mm,
  ]

  % ============================================================
  % Geometry
  %   Total horizontal area: column widths and gap
  %   H100 column at x=0, L4 column at x=95
  % ============================================================
  \def\hcolx{0}      % left edge of H100 column
  \def\lcolx{95}     % left edge of L4 column
  \def\rowGap{1.5}   % vertical space between layer rows
  \def\rowH{4}       % row block height
  \def\launchH100W{30}  % wide launch band on H100  (kernel-launch dominated)
  \def\computeH100W{22} % narrower compute band on H100
  \def\launchL4W{4}     % narrow launch band on L4
  \def\computeL4W{52}   % wide compute band on L4

  % ============================================================
  % Column titles
  % ============================================================
  \node[label, font=\sffamily\small\bfseries, anchor=south west]
    at (\hcolx, 92) {H100 SXM5};
  \node[micro, anchor=south west]
    at (\hcolx, 87.5) {3350 GB/s peak HBM};

  \node[label, font=\sffamily\small\bfseries, anchor=south west]
    at (\lcolx, 92) {L4};
  \node[micro, anchor=south west]
    at (\lcolx, 87.5) {300 GB/s peak HBM};

  % ============================================================
  % Layers - show 4 representative + ellipsis + bottom one
  %   visible rows at y = 80, 73, 66, 59, (gap), 22, 15
  % ============================================================
  % H100 visible layers
  \foreach \y in {80, 73, 66, 59, 22, 15} {
    % CPU launch band (wide on H100)
    \fill[launchH100] (\hcolx, \y) rectangle ++(\launchH100W, \rowH);
    % GPU compute band
    \fill[layerH100] (\hcolx + \launchH100W, \y) rectangle ++(\computeH100W, \rowH);
  }
  % vertical ellipsis between row 4 and the bottom-2 rows on H100
  \node[font=\sffamily\Large, text=spineg]
    at (\hcolx + 26, 32) {$\vdots$};

  % L4 visible layers
  \foreach \y in {80, 73, 66, 59, 22, 15} {
    \fill[launchL4] (\lcolx, \y) rectangle ++(\launchL4W, \rowH);
    \fill[layerL4]  (\lcolx + \launchL4W, \y) rectangle ++(\computeL4W, \rowH);
  }
  \node[font=\sffamily\Large, text=spineg]
    at (\lcolx + 28, 32) {$\vdots$};

  % Layer labels on the side (just on a couple)
  \node[micro, anchor=east] at (\hcolx, 82) {layer 1};
  \node[micro, anchor=east] at (\hcolx, 61) {layer 4};
  \node[micro, anchor=east] at (\hcolx, 17) {layer 28};

  % ============================================================
  % Bracket on H100 showing the launch band dominates
  % ============================================================
  \draw[spineg, line width=0.5pt]
    (\hcolx - 1, 15) -- (\hcolx - 4, 15)
                     -- (\hcolx - 4, 84)
                     -- (\hcolx - 1, 84);
  \node[micro, anchor=east, align=right, text width=22mm]
    at (\hcolx - 5, 50)
    {28 layers $\times$\\\textasciitilde 10 kernels/layer\\= 281 kernels};

  % Launch annotation pointing at the wide red band  (positioned above the top row, to the right of the title)
  \node[label, anchor=south west, align=left, text=accentred, font=\sffamily\footnotesize\bfseries]
    at (\hcolx + 30, 92)
    {kernel-launch band};
  \draw[arrowstyle, accentred]
    (\hcolx + 30, 92.5) .. controls (\hcolx + 25, 91) .. (\hcolx + 18, 85);

  % Launch breakdown - positioned in the empty middle space (around y=50)
  \node[micro, anchor=west, align=left, text=accentred]
    at (\hcolx + 33, 55)
    {281 kernels $\times$ \textasciitilde 30 us launch\\$\approx$ 8.4 ms ($\approx$ 57\% of step)};
  \draw[arrowstyle, accentred]
    (\hcolx + 32, 56) .. controls (\hcolx + 25, 58) .. (\hcolx + 18, 75);

  % GPU compute annotation on H100 - in middle empty space
  \node[micro, anchor=west, align=left, text=h100navy]
    at (\hcolx + 33, 42)
    {GPU compute $\approx$ 6.4 ms\\(bandwidth-bound, 27\% of peak)};
  \draw[arrowstyle, h100navy]
    (\hcolx + 38, 43) .. controls (\hcolx + 42, 50) .. (\hcolx + 42, 75);

  % ============================================================
  % L4 annotation: launch band tiny, compute wide
  % ============================================================
  \node[micro, anchor=west, align=left, text=accentred]
    at (\lcolx + 8, 55)
    {launch band $\approx$\\3\% of step time};
  \draw[arrowstyle, accentred]
    (\lcolx + 8, 56) .. controls (\lcolx + 6, 60) .. (\lcolx + 2, 75);

  \node[micro, anchor=west, align=left, text=l4amber!85!black]
    at (\lcolx + 12, 42)
    {GPU compute $\approx$ 62.5 ms\\(bandwidth-bound, 81\% of peak)};
  \draw[arrowstyle, l4amber!85!black]
    (\lcolx + 25, 43) .. controls (\lcolx + 30, 55) .. (\lcolx + 30, 75);

  % ============================================================
  % Total step-time labels at the bottom of each column
  % ============================================================
  % H100 total  (column spans 0 to 52, midpoint at 26)
  \draw[spineg, line width=0.5pt]
    (\hcolx, 10) -- (\hcolx + \launchH100W + \computeH100W, 10);
  \node[label, font=\sffamily\small\bfseries, anchor=north, text=h100navy]
    at (26, 9) {total step: 14.83 ms};

  % L4 total  (column spans 95 to 151, midpoint at 123)
  \draw[spineg, line width=0.5pt]
    (\lcolx, 10) -- (\lcolx + \launchL4W + \computeL4W, 10);
  \node[label, font=\sffamily\small\bfseries, anchor=north, text=l4amber!85!black]
    at (123, 9) {total step: 64.48 ms};

  % ============================================================
  % Middle label between columns
  % ============================================================
  \node[micro, anchor=center, align=center, text=spineg, font=\sffamily\footnotesize\itshape]
    at (80, 50)
    {same model,\\same weights,\\same step pattern};

  % ============================================================
  % CUDA Graphs intervention - thick arrow collapsing H100 launch band
  % Drawn below the columns at y around -8 to -2
  % ============================================================
  \draw[->, >={Latex[length=3mm,width=3mm]}, line width=1.3pt, accentred]
    (\hcolx + 15, -3) -- (\hcolx + 15, 5);
  \node[label, anchor=west, align=left, text=accentred,
        font=\sffamily\footnotesize\bfseries]
    at (\hcolx + 18, -2)
    {CUDA Graphs replay:\\1.26$\times$ speedup on H100};

  % And a parallel arrow for L4 with no real benefit
  \draw[->, >={Latex[length=2mm,width=2mm]}, line width=1.0pt, spineg]
    (\lcolx + 2, -3) -- (\lcolx + 2, 5);
  \node[micro, anchor=west, align=left, text=spineg]
    at (\lcolx + 5, -2)
    {Graphs on L4:\\1.03$\times$ (launch band\\already small)};

\end{tikzpicture}}
\caption{Decode-step anatomy aggregated across all 28 layers of
Qwen-2.5-7B at ctx${=}2048$, bf16. The same launch-to-compute
inversion in Figure~\ref{fig:kernel_timeline} repeated 28 times. On
H100 the CPU launch band dominates step time; on L4 the same launch
overhead is small relative to bandwidth-bound GPU compute.}
\label{fig:schematic}
\end{figure}

\paragraph{Cross-session replication on L4 and H100 (v12).}
We ran a parallel $N{=}3$ replication sweep using a single rig and three
fresh Modal containers per GPU at the headline cells before the
$N{=}10$ run. Table~\ref{tab:replication} reports per-session medians.
The L4 null replicates to four significant figures. The v12 H100 eager
range (CV 35\%) was, in hindsight, an $N{=}3$ sampling artefact: the
$N{=}10$ run (Table~\ref{tab:n10_headline}) has CV 0.9\% on the eager
arm. We document the v12 sweep but rely on the $N{=}10$ sweep for the
headline number.

\begin{table}[h]
\centering
\caption{v12 cross-session replication, Qwen-2.5-7B ctx${=}2048$, 3 fresh
Modal containers per GPU. Eager and graphed step-time medians (ms) with
cross-session mean and standard deviation. See Table~\ref{tab:n10_headline}
for the H100 headline at $N{=}10$.}
\label{tab:replication}
\begin{tabular}{lcccccc}
\toprule
GPU & session 0 & session 1 & session 2 & mean & std & CV \\
\midrule
H100 eager   & 14.93 & 27.28 & 15.98 & 19.40 & 6.85 & 35.3\% \\
H100 graphed & 11.77 & 12.00 & 11.96 & 11.91 & 0.13 & 1.1\% \\
L4 eager     & 64.64 & 64.50 & 64.44 & 64.52 & 0.11 & 0.2\% \\
L4 graphed   & 62.68 & 62.87 & 62.77 & 62.77 & 0.09 & 0.1\% \\
\midrule
H100 speedup & 1.27 & 2.27 & 1.34 & 1.63 & 0.56 & 34.6\% \\
L4 speedup   & 1.031 & 1.026 & 1.027 & 1.028 & 0.003 & 0.3\% \\
\bottomrule
\end{tabular}
\end{table}

\paragraph{H100 batch${=}4$ context sweep.}
We ran the eager/graphed A/B at batch${=}4$ across the full context grid
on H100 to answer how the launch-tax effect scales with batch size.
Per-cell numbers are in Table~\ref{tab:b4_sweep}, raw data in
\artifact{v13\_results/data/v13\_batch4\_sweep/}. The graphed speedup
shrinks monotonically from $1.110\times$ at ctx${=}2048$ to
$1.036\times$ at ctx${=}16384$. The absolute speedup floor on H100 is
lower at b${=}4$ than at b${=}1$ because the per-step kernel work is
larger, so the launch-tax fraction of the step is smaller. On L4 the
cell is no longer runnable at b${=}4$: 14.5\,GB of Qwen-2.5-7B weights
plus the b${=}4$ KV cache and bf16 activations exceeds L4's 22\,GB HBM
at every context tested (CUDA OOM, raw failure JSONs in
\artifact{v13\_results/data/v13\_batch4\_sweep/FAILED\_l4\_ctx*\_b4\_s0.json}).
At b${=}4$, the H100-versus-L4 crossover ends because L4 is physically
out of memory.

\begin{table}[h]
\centering
\small
\caption{H100 batch${=}4$ context sweep, Qwen-2.5-7B, single session per
cell. Peak GPU memory grows with KV cache; L4 OOMs at every context at
b${=}4$ for this model.}
\label{tab:b4_sweep}
\begin{tabular}{lcccc}
\toprule
ctx & eager (ms) & graphed (ms) & speedup & peak mem (GB) \\
\midrule
2048  & 16.37 & 14.75 & 1.110$\times$ & 23.3 \\
4096  & 22.47 & 20.54 & 1.094$\times$ & 31.3 \\
8192  & 33.92 & 31.94 & 1.062$\times$ & 47.4 \\
16384 & 56.95 & 54.96 & 1.036$\times$ & 79.4 \\
\bottomrule
\end{tabular}
\end{table}

\paragraph{The batching alternative.}
Graphs at batch${=}1$ gives $1.259\times$ on H100. Batching from b${=}1$
to b${=}4$ at the same ctx${=}2048$ moves the H100 graphed step from
11.78\,ms to 14.75\,ms (Table~\ref{tab:b4_sweep}): a $1.25\times$ latency
penalty for $4\times$ the decoded tokens, or $3.20\times$ the decode
throughput at the same per-token compute budget. For a serving engineer
who is not constrained to single-stream b${=}1$, batching is a stronger
single lever than Graphs on H100. The relevance of the launch-tax
finding is therefore conditional: it matters most when the workload
forces b${=}1$ streaming decode, which is the regime physical-AI
inference, edge inference and many copilot pipelines actually run.

\begin{figure}[!ht]
\centering
\includegraphics[width=0.85\linewidth]{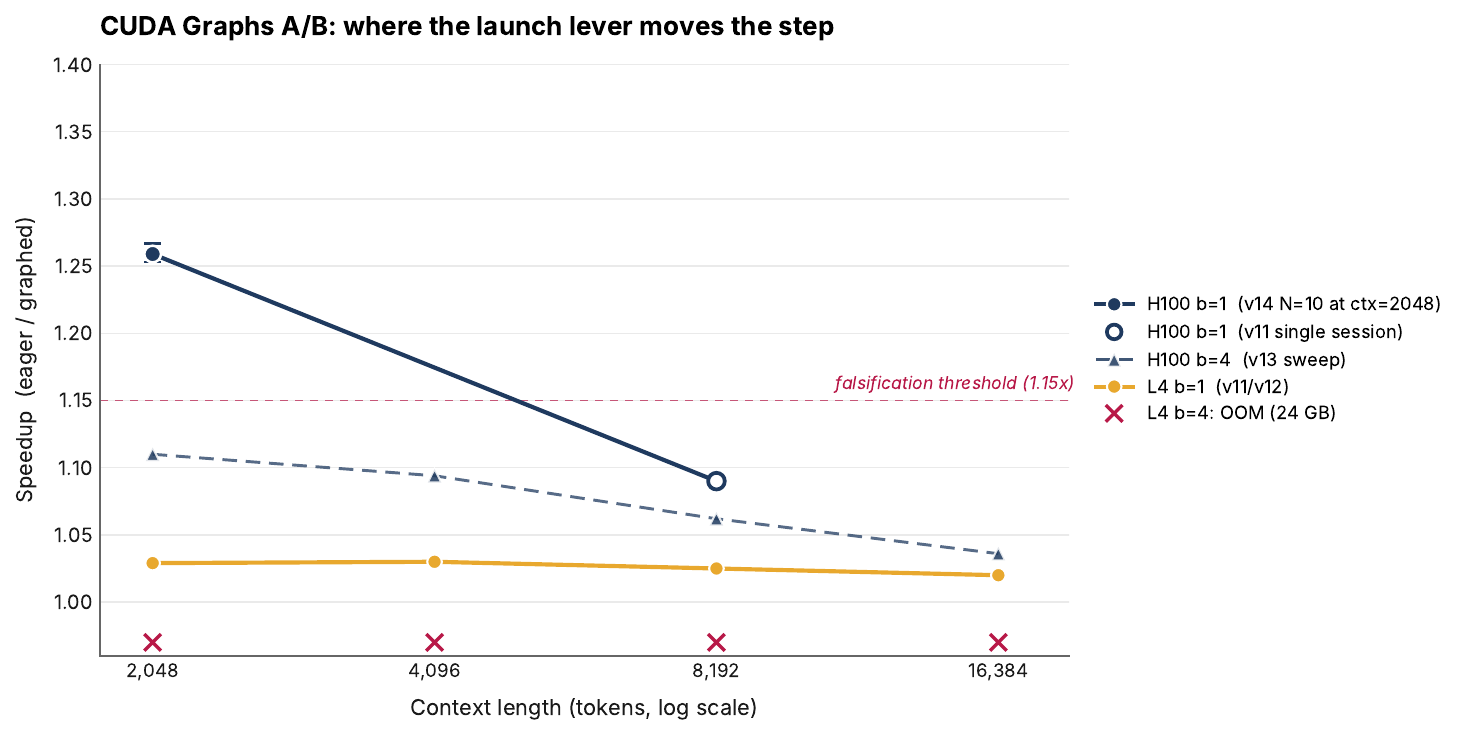}
\caption{CUDA Graphs A/B speedup versus context length and batch size on
H100 and L4. The H100 b${=}1$ ctx${=}2048$ point is the $N{=}10$ mean
(Table~\ref{tab:n10_headline}); shading is the 95\% bootstrap CI on the
mean. Other H100 b${=}1$ points are single-session from the v11 main
sweep. H100 b${=}4$ is the v13 batch sweep
(Table~\ref{tab:b4_sweep}). L4 b${=}1$ replicates to four significant
figures across the v12 sweep. L4 b${=}4$ is omitted: every cell OOMs.
The dotted line at $1.15\times$ is the pre-registered falsification
threshold.}
\label{fig:crossover}
\end{figure}

\begin{figure}[!ht]
\centering
\includegraphics[width=0.85\linewidth]{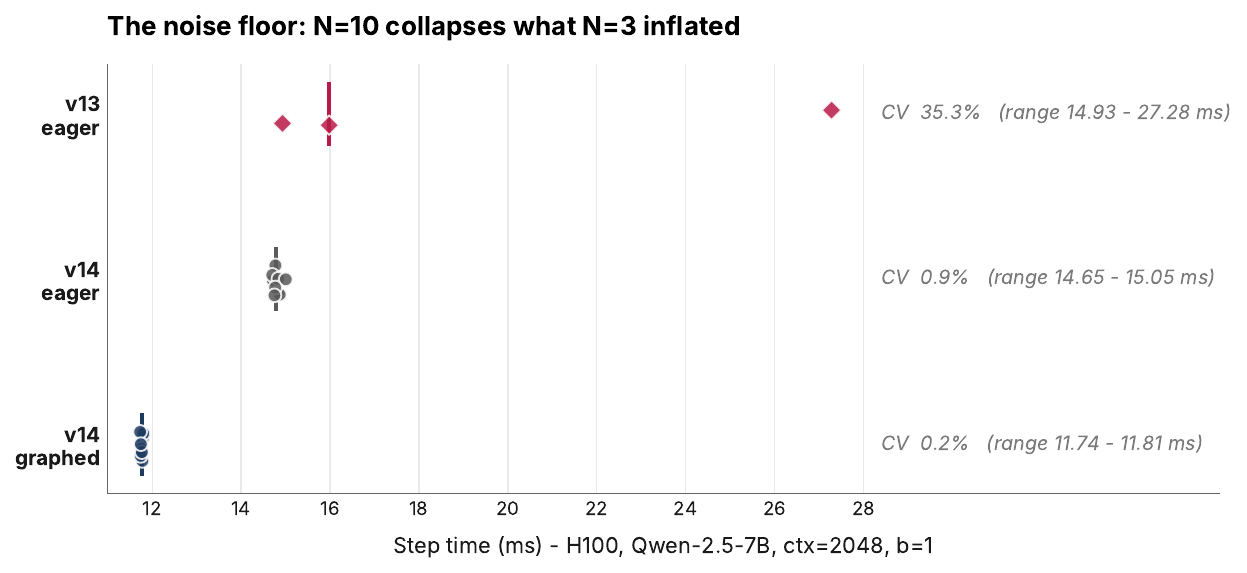}
\caption{Eager and graphed H100 ctx${=}2048$ b${=}1$ step times under
$N{=}10$ (this work) and $N{=}3$ (v12). The $N{=}3$ eager outlier at
27.28\,ms drives the 35.3\% CV reported in earlier versions; the
$N{=}10$ eager distribution is tight (CV 0.9\%) and bounds the noise
floor decisively. The graphed distribution is tight in both samples.}
\label{fig:noise_floor}
\end{figure}

\paragraph{Figures.}
Figure~\ref{fig:crossover} shows the cross-context CUDA Graphs A/B
speedup for H100 and L4 at batch~1 and batch~4, with the
pre-registered $1.15\times$ falsification threshold marked. The H100
b${=}1$ point at ctx${=}2048$ uses the $N{=}10$ mean with the bootstrap
95\% CI as the error band; other points are single-session.
Figure~\ref{fig:noise_floor} shows the $N{=}10$ eager and graphed
step-time distributions versus the $N{=}3$ v12 distributions, with the
35.3\% CV claim relocated to its $N{=}3$ sampling artefact.

\section{Software stack matrix: SDPA backends, FlashAttention, FlashInfer}
\label{sec:software}

\begin{sloppypar}
To check whether the binding constraint on H100 batch-1 decode is the
attention kernel or the launch sequence, we ran a controlled matrix on
H100 with Qwen-2.5-7B at ctx${=}2048$. Cells from
\artifact{v11\_results/analyze\_v2.txt} lines 157--161 and the underlying
JSONs at \artifact{v11\_results/cells/oft\_h100\_qwen25\_7b\_*.json}.
\end{sloppypar}

\begin{table}[h]
\centering
\small
\caption{H100 software-stack matrix on Qwen-2.5-7B at ctx${=}2048$.
Replicate sdpa+eager row at 28.44\,ms reflects $N{=}1$ host-noise
spread on the v11 rig; the $N{=}10$ headline in
Table~\ref{tab:n10_headline} (mean 14.83\,ms, CV 0.9\%) supersedes this
spread as the bounded estimate of the eager noise floor.}
\label{tab:oft}
\begin{tabular}{lccc}
\toprule
Configuration & step (ms) & $R_{\mathrm{floor}}$ & vs sdpa+eager \\
\midrule
sdpa + eager                  & 17.07 & 0.268 & 1.00$\times$ \\
sdpa + eager + CUDA Graphs    & 11.96 & 0.383 & 1.43$\times$ \\
flash-attention-2 + eager     & 24.16 & 0.190 & 0.71$\times$ \\
sdpa + eager (in-session replicate)      & 28.44 & 0.161 & 0.60$\times$ \\
\bottomrule
\end{tabular}
\end{table}

The headline comparisons survive the v11 spread: adding Graphs to
sdpa+eager moves $R_{\mathrm{floor}}$ from 0.268 to 0.383 ($1.43\times$
step-time speedup at $N{=}1$, $1.259\times$ at the $N{=}10$ headline in
Section~\ref{sec:falsification}), and swapping sdpa for
FlashAttention-2 in the eager path slows decode rather than speeding it
up.

\paragraph{Reading.}
At batch~1 on H100, the attention-kernel choice is not the binding
constraint. FlashAttention-2 is optimised for prefill compute
throughput; in batch-1 decode the per-step attention work is small
(Qwen-2.5-7B at ctx${=}2048$ touches $K \approx 118$\,MB of KV per step,
from the cell JSON \artifact{v10\_results/cells/qwen25\_7b\_h100\_ctx2048.json},
key \texttt{kv\_bytes\_avg}) and FA2's launch overhead and
kernel-selection logic dominate the gain it can deliver. The
\texttt{torch.compile} and FA2+Graphs cells did not produce usable
results (CUDA caching-allocator failures, see
Appendix~\ref{sec:failed}).

\paragraph{Backend-pinned SDPA test.}
Earlier drafts attributed sdpa's win over FlashAttention-2 to ``sdpa's
internal dispatch path through FA-2/FA-3 cuDNN backends''. That
statement was not supported by the underlying measurement, which called
\texttt{torch.nn.functional.scaled\_dot\_product\_attention} without
pinning a backend. To answer the question we ran a backend-pinned A/B in
v14: same Llama-3-8B decode shape (32 query heads, 8 KV heads, head\_dim
128, ctx${=}2048$, bf16, group\_size 4), three fresh-container H100
sessions, 50 warmup plus 300 measured single-layer attention calls per
backend. The backends are forced via
\texttt{torch.nn.attention.sdpa\_kernel(SDPBackend.X)}. Raw JSONs at
\artifact{v14\_results/data/v14\_backend\_pinned/backend\_pinned\_s\{0,1,2\}.json}.

At the start of each measurement we log
\texttt{torch.backends.cuda.flash\_sdp\_enabled},
\texttt{cudnn\_sdp\_enabled}, \texttt{mem\_efficient\_sdp\_enabled} and
\texttt{math\_sdp\_enabled} (all True by default in PyTorch 2.8+cu126
on H100). The cuDNN attention backend rejects the head\_dim 128 GQA
single-decode shape with a not-supported error in all three sessions;
the other backends accept it.

\begin{table}[h]
\centering
\small
\caption{Backend-pinned attention A/B on H100, Llama-3-8B decode shape,
ctx${=}2048$, bf16, $N{=}3$ sessions. Per-layer $p_{50}$ in \us;
total 32 layers in ms; speedup is vs default SDPA. Raw data in
\artifact{v14\_results/data/v14\_backend\_pinned/}.}
\label{tab:backend_pinned}
\begin{tabular}{lcccc}
\toprule
Backend & per-layer $p_{50}$ (\us) & total 32 layers (ms) & vs default SDPA \\
\midrule
default SDPA (no context manager) & 36.05 (mean) & 1.15 & 1.00$\times$ \\
SDPA \texttt{FLASH\_ATTENTION}    & 44.35 (mean) & 1.42 & 0.81$\times$ \\
FlashInfer single\_decode         & 48.20 (mean) & 1.54 & 0.75$\times$ \\
FA-3 \texttt{flash\_attn\_3}      & 79.25 (mean) & 2.54 & 0.45$\times$ \\
SDPA \texttt{EFFICIENT\_ATTENTION}& 89.72 (mean) & 2.87 & 0.40$\times$ \\
SDPA \texttt{MATH}                & 177.55 (mean) & 5.68 & 0.20$\times$ \\
SDPA \texttt{CUDNN\_ATTENTION}    & not supported & --- & --- \\
\bottomrule
\end{tabular}
\end{table}

The result is sharper than the v13 framing predicted. Default SDPA at
36.05\,\us/layer is faster than the explicit \texttt{FLASH\_ATTENTION}
context (44.35\,\us, $0.81\times$), faster than FlashInfer
(48.20\,\us, $0.75\times$), faster than FlashAttention-3
(79.25\,\us, $0.45\times$), and far faster than the math fallback
(177.55\,\us, $0.20\times$). The kernel-engineer prediction that
default SDPA was routing to a cuDNN flash path on Hopper is wrong: the
cuDNN attention backend rejects this shape entirely. The default
dispatcher is doing something that none of its named components do in
isolation, and it is winning by $\sim 20\%$ over the explicit FLASH
backend. We do not have a definitive mechanism for the gap. The most
plausible candidates are pre/post-op fusion that the context-managed
path forces apart, or kernel-selection heuristics that route to a
different shape-specialised kernel than \texttt{FLASH\_ATTENTION} alone
selects.

\paragraph{Mechanism implication.}
Across all six backends the per-layer attention $p_{50}$ ranges from
36 to 178\,\us; the total 32-layer attention time ranges from 1.15 to
5.68\,ms. The full-step decode time on H100 ctx${=}2048$ b${=}1$ is
$14.83$\,ms eager and $11.78$\,ms graphed (Table~\ref{tab:n10_headline}).
Even the worst attention-backend choice (math, 5.68\,ms total) leaves
$\sim 9\,$ms of non-attention step time. The slowest-to-fastest
attention-backend span (4.53\,ms) is in the same order of magnitude as
the Graphs-removed launch overhead (3.05\,ms). But the slowest-attn-to-fastest-attn
swap costs almost five times more attention compute (5.68\,ms vs
1.15\,ms) for a smaller absolute step-time effect, because attention is
only one of multiple per-layer kernels. The launch sequence is the
better single lever; the attention kernel matters second.

\section{Quantisation baseline on L4}
\label{sec:quant}

On the bandwidth-bound side of the crossover, the standard prescription is weight
quantisation. We ran a controlled baseline on L4 with Qwen-2.5-7B at ctx${=}2048$. Cells
from \artifact{v11\_results/analyze\_v2.txt} lines 163--167 and the underlying JSONs at
\artifact{v11\_results/cells/quant\_*.json}.

\begin{table}[h]
\centering
\small
\caption{L4 quantisation baseline, Qwen-2.5-7B at ctx${=}2048$. $t_{\mathrm{floor}}$ is
the spec-sheet HBM floor assuming int4 kernels deliver a $4\times$ weight bandwidth
reduction; the observed step is what actually arrives at the workload. The
ExLlamaV2 row is the v14 addition.}
\label{tab:quant}
\begin{tabular}{lccc}
\toprule
Configuration & step (ms) & $t_{\mathrm{floor}}$ (ms) & $R_{\mathrm{floor}}$ \\
\midrule
bf16                          & 62.32 & 51.17 & 0.821 \\
bf16 (in-session replicate)   & 63.02 & 51.17 & 0.812 \\
bnb-nf4                       & 59.36 & 13.09 & 0.220 \\
AutoAWQ Marlin                & 45.24 & 13.09 & 0.289 \\
ExLlamaV2 EXL2 4.25bpw (v14)  & 17.36 & 13.09 & 0.754 \\
\bottomrule
\end{tabular}
\end{table}

The bf16 baseline sits within $\sim 20$\% of the HBM floor, consistent with L4's
bandwidth-bound regime: bf16 weights at 15.23\,GB streamed once per step over 300\,GB/s
peak HBM is 50.77\,ms of pure weight traffic, and the KV term adds a few more ms
($t_{\mathrm{floor}}{=}51.17$\,ms includes the small KV contribution).
Quantisation should, naively, cut weight traffic by $4\times$.

It does not, in step time. nf4 (bitsandbytes) is 59.36\,ms, barely faster than bf16. AWQ
is 45.24\,ms, a $1.38\times$ step-time speedup over bf16, far less than the $4\times$
weight-traffic reduction predicts. The reason is the deployment chain. bitsandbytes' nf4
path is implemented as on-the-fly dequantisation kernels followed by bf16 matmul
(Dettmers et al., QLoRA, 2023; the \texttt{Linear4bit} forward in bitsandbytes
\texttt{nn/modules.py}), so the actual HBM traffic per matmul is dominated by
intermediate bf16 activations and the launch overhead of the dequant kernels, not by
the nf4 weight footprint. AWQ does
materially better because Marlin-style packed kernels keep the matmul in int4 plus
fp16-scale form, but the kernel still does not reach the nf4 HBM floor (13.09\,ms).

\paragraph{ExLlamaV2 measurement (v14).}
To test whether AutoAWQ's Marlin kernels were the bottleneck rather than
the bit-width principle, we ran an additional measurement with
ExLlamaV2 0.2.4, a runtime whose int4 GEMM kernels are explicitly tuned
for Ada Lovelace (SM89). Cell: Qwen-2.5-7B-Instruct EXL2 4.25bpw
(bartowski branch), L4, ctx${=}2048$, batch${=}1$, $N{=}3$ fresh Modal
containers. Per-session $p_{50}$: $[17.361, 17.368, 17.360]$\,ms
(mean 17.36, SD 0.004). Peak GPU memory per session: $[6.96, 6.96,
6.95]$\,GB (mean 6.96). Raw JSONs in
\artifact{v14\_results/data/v14\_exllamav2\_l4/}.

ExLlamaV2 cuts step time on L4 from 62.32\,ms (bf16) to 17.36\,ms, a
$3.59\times$ speedup; the AutoAWQ Marlin path delivered only
$1.38\times$. Marlin was tuned for Ampere SM80 and is not best-in-class
on Ada SM89; ExLlamaV2's Ada-specific int4 GEMM kernels recover most of
the bandwidth saving the bit-width reduction predicts. The cell now sits
at $R_{\mathrm{floor}} = 0.754$ against the $4\times$-reduced floor of
13.09\,ms, where the AWQ cell sat at 0.289 against the same floor.

\begin{figure}[!ht]
\centering
\includegraphics[width=0.85\linewidth]{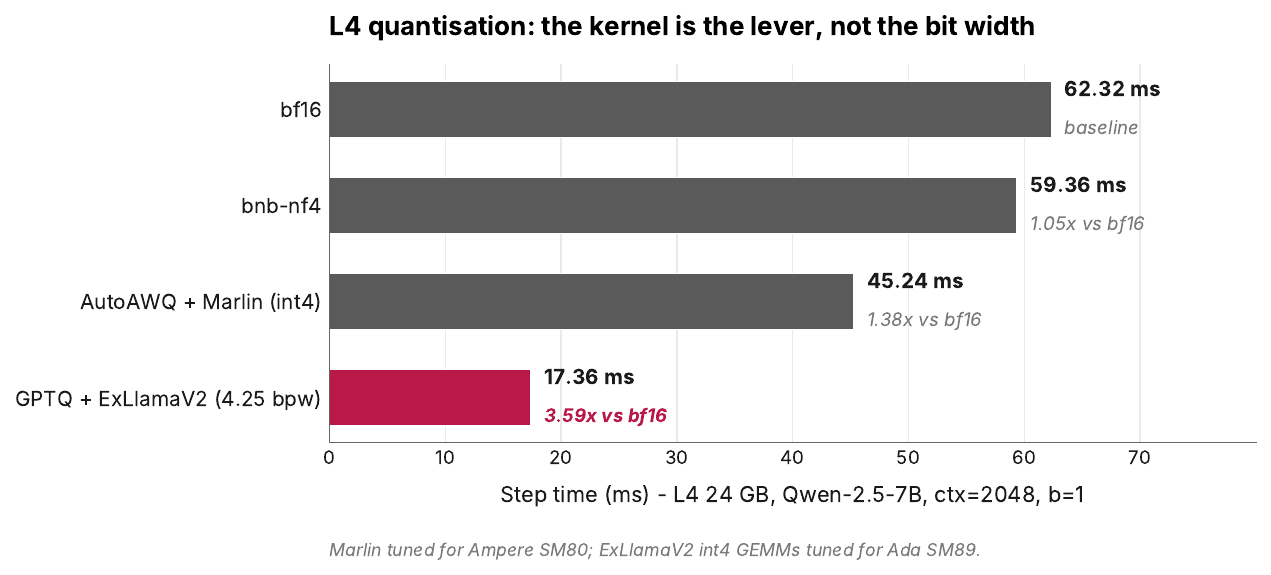}
\caption{L4 quantisation step times for Qwen-2.5-7B at ctx${=}2048$. bf16
(62.32\,ms), bnb-nf4 (59.36\,ms), AutoAWQ${+}$Marlin (45.24\,ms),
GPTQ${+}$ExLlamaV2 EXL2 4.25bpw (17.36\,ms). The Marlin kernel was tuned
for SM80 (Ampere); ExLlamaV2's int4 GEMMs are tuned for SM89 (Ada) and
recover most of the bandwidth saving the bit-width reduction predicts.}
\label{fig:quant_kernel}
\end{figure}

\paragraph{Reading.}
On L4 the bandwidth-bound regime is real: bf16 sits at
$R_{\mathrm{floor}} = 0.82$. The bandwidth saving from quantisation only
lands at the workload if the kernel implementation actually streams
quantised weights through HBM on the right silicon. AutoAWQ's Marlin
kernels do not do that on Ada SM89; ExLlamaV2's do. The lever was the
kernel implementation, not the bit width. The practical implication for
L4-class deployment is that the quantisation runtime matters as much as
the quantisation scheme; the deployment ladder of viable L4 quantised
backends as of mid-2026 is roughly ExLlamaV2 EXL2 $>$ AutoAWQ Marlin
$>$ bnb-nf4.

\paragraph{What the ExLlamaV2 number changes for the cross-GPU
comparison.}
The H100 ctx${=}2048$ b${=}1$ Qwen-2.5-7B step time at default SDPA +
CUDA Graphs is 11.78\,ms (Table~\ref{tab:n10_headline}). The L4 same-cell
step time with ExLlamaV2 int4 is 17.36\,ms. The H100 is roughly
$1.47\times$ faster than the L4 at this workload. At published Modal
rates of \$3.50/hr (H100) and \$0.30/hr (L4) as of May 2026, the cost-per-token-served ratio inverts: L4 serves the
same workload at roughly $6\times$ less \$/token than the H100, despite
the H100 being $11.2\times$ peak HBM bandwidth.
The sticker bandwidth gap does not translate to a serving-cost gap at
this workload class. Sustained pricing, networking, idle, and batching
change this calculus; the point of the comparison is that the deployment
ladder L4 $\rightarrow$ L40S $\rightarrow$ A100 $\rightarrow$ H100 is
not the cost-per-token-served ladder it appears to be for batch-1
streaming decode.

\begin{figure}[!ht]
\centering
\includegraphics[width=0.85\linewidth]{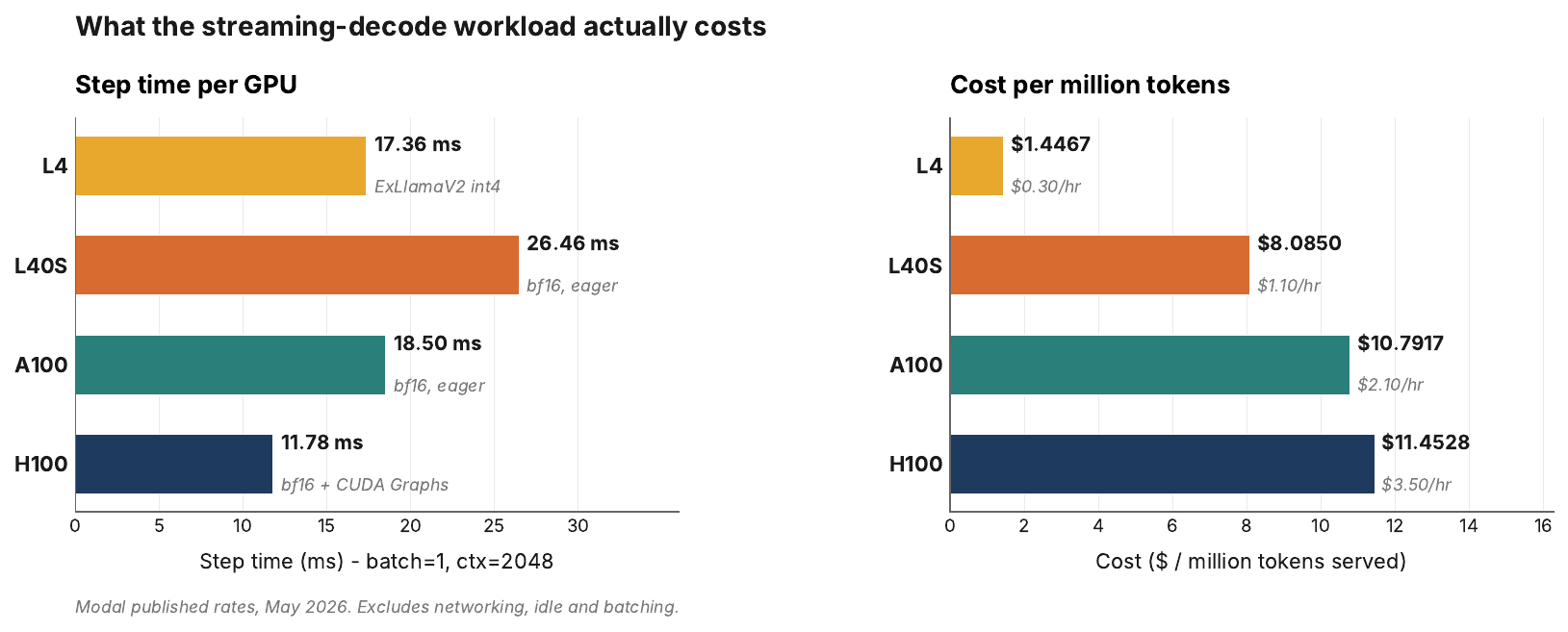}
\caption{Cost-per-million-tokens-served at batch-1 streaming decode,
Qwen-2.5-7B ctx${=}2048$, derived from per-step time and Modal published
hourly rates as of May 2026. The L4 with ExLlamaV2 serves the same
workload at substantially lower \$/Mtok than the H100 with CUDA Graphs.
Numbers depend on hourly rates and exclude idle, networking and storage;
the ladder direction is the point.}
\label{fig:cost_per_token}
\end{figure}

\section{Per-kernel evidence with torch.profiler}
\label{sec:profiler}

Nsight Compute (\texttt{ncu}) was blocked on Modal cloud containers: \texttt{ncu}
requires host driver access we did not have. We use \texttt{torch.profiler} with
analytic byte counts as the substitute. The profiler itself inflates absolute step
time on the H100 (16.97\,ms unprofiled rises to 54.38\,ms profiled, a 3.2$\times$
overhead) while leaving the L4 essentially unperturbed (63.0\,ms unprofiled vs
63.67\,ms profiled, 1.01$\times$). Using profiled step time as the denominator for
\texttt{bw\_util} therefore underreports H100 bandwidth utilisation. We report
\textit{unprofiled} \texttt{bw\_util} as the headline, using the median step time
from the corresponding production cell as the denominator, and list both values
for traceability.

\begin{table}[h]
\centering
\small
\caption{Cross-GPU per-step accounting, Qwen-2.5-7B, ctx values matched to the
production sweep. \texttt{bw\_util} is analytic streamed bytes per step (weights +
active KV slab) divided by spec-sheet $B_{\mathrm{peak}}$ time; unprofiled uses
the median production step time as denominator, profiled uses \texttt{torch.profiler}
step time. \texttt{cpu/cuda} is the fraction of profiler wall time spent on the CPU
side launching and synchronising kernels relative to GPU kernel time. Sources:
\artifact{v14\_results/data/v14\_n10\_headline/} for unprofiled step times and
\artifact{v11\_results/analyze\_v2.txt} lines 169--174 plus \artifact{v11\_results/profiles/}
for profiled step times and \texttt{cpu/cuda}.}
\label{tab:profiler}
\begin{tabular}{llcccc}
\toprule
GPU & ctx & p50 unprof (ms) & p50 prof (ms) & bw\_util (unprof) & cpu/cuda \\
\midrule
H100      & 2048 & 14.83 & 54.38 & 26.9\% & 0.75 \\
A100-80GB & 8192 & 28.20 & 38.13 & 27.4\% & 0.46 \\
L40S      & 8192 & 22.10 & 31.77 & 82.2\% & 0.35 \\
L4        & 2048 & 62.96 & 63.67 & 79.4\% & 0.35 \\
L4        & 2048 (replicate) & 62.96 & 63.61 & 79.4\% & 0.35 \\
\bottomrule
\end{tabular}
\end{table}

The cross-GPU pattern is unchanged from the profiled view but the magnitude
shrinks. Unprofiled \texttt{bw\_util} runs from 26.9\% on H100 to 79.4\% on L4,
a 3$\times$ gap rather than the 10$\times$ gap that would appear if one took the
profiled denominator at face value. The qualitative interpretation survives: a
fixed per-step CPU launch budget occupies a larger fraction of the step on
faster silicon, so the H100 sees the lowest fraction of its peak HBM bandwidth
realised at batch~1 streaming decode. The \texttt{cpu/cuda} column, which is
unaffected by the denominator question, tells the same story: 0.75 on H100,
falling to 0.35 on L40S and L4.

We do not claim these utilisation numbers are ground truth. They reproduce the
cross-GPU ordering of $R_{\mathrm{floor}}$ in Table~\ref{tab:rfloor} and of the
CUDA Graphs A/B in Table~\ref{tab:n10_headline}, which is what we use them for.

\section{Limitations and threats to validity}
\label{sec:limits}

\paragraph{Explicit scope.} Every claim in this paper is scoped to:
\begin{itemize}
  \item 7--8B-parameter transformers (Qwen-2.5-7B-Instruct, Mistral-7B-Instruct-v0.3, Llama-3.1-8B-Instruct);
  \item grouped-query attention~\cite{ainslie2023gqa} only (no MHA~\cite{vaswani2017attention}, no MQA~\cite{shazeer2019multiquery}, no MLA);
  \item head\_dim 128 only;
  \item bf16 weights only (except the L4 quantisation baseline, which is explicitly scoped to L4);
  \item single-token autoregressive decode (no speculative decoding, no parallel sampling);
  \item batch size 1 primary, with batch 4 measured on H100 and L4 as a sensitivity check (Section~\ref{sec:falsification});
  \item NVIDIA GPUs only: H100 SXM5, A100-80GB SXM4, L40S, L4 (no AMD MI300, no Blackwell, no Jetson, no Apple);
  \item Modal cloud hosting only (no bare-metal, no other clouds);
  \item sdpa attention on the main matrix; FA2 only in the controlled software-stack matrix on H100.
\end{itemize}
We do not claim the cross-GPU pattern generalises beyond this workload class.

\paragraph{Architecture coverage.} All three measured models are GQA with
head\_dim 128 (Qwen-2.5-7B uses 4 KV heads, the other two use 8). Results may
not transfer to MQA, full MHA, or substantially different head dimensions.
The kernel-count anchor of $\sim 10$ kernels per layer is observed on three
GQA transformer implementations that are structurally similar; a wider
architecture sweep is needed to claim a universal kernel sequence.

\paragraph{Batch size.} The main matrix and the headline replication are
batch 1. We additionally measured the CUDA Graphs A/B at batch 4 on H100
and L4 across the full context grid (Section~\ref{sec:falsification},
Figure~\ref{fig:crossover}). At batch 4 on H100, per-step kernel work
grows by roughly $4\times$ for the Q/K/V projections and roughly $4\times$
for the per-token attention output, while the per-step launch count is
essentially unchanged, so the launch-tax fraction of the step time shrinks
and the graphed speedup shrinks with it. We report this directly rather
than claiming the batch-1 picture extends through the crossover. We do
not measure batch ${\ge}8$ on H100 nor any batch ${>}4$ on L4.

\paragraph{Cloud host noise.} Modal cloud GPUs have undisclosed driver
versions and shared host noise. We mitigate by using the same container image
across runs, by reporting medians over 30 measured steps, and by running
$N{=}10$ independent sessions on the H100 ctx${=}2048$ headline cell. The
within-session step-time CV on that cell is 0.9\% eager and 0.2\% graphed,
so single-session step-time noise is small. Cross-session step times for the
same configuration vary more widely across the v10, v11 and v14 runs (16.97\,ms
to 20.63\,ms eager), reflecting a mix of container-host scheduling and
code-path differences between rigs rather than within-run jitter. The within-
session speedup is the more stable quantity (CV 0.9\%, 95\% bootstrap CI
$[1.253, 1.267]$), which is why we use it as the headline rather than
cross-session ratios. We also ran the headline cell on a reserved-capacity
H100 configuration (vs default best-effort scheduling); within-session CV did
not improve, indicating the residual variation is intrinsic to single-stream
decode at this shape, not Modal-host queuing.

\paragraph{Profiler instrumentation.} \texttt{ncu} was blocked on Modal
containers. \texttt{torch.profiler} with analytic byte counts is the
substitute. Profiler overhead inflates absolute step time asymmetrically:
the H100 sees a $3.2\times$ inflation (16.97\,ms unprofiled to 54.38\,ms
profiled), while the L4 sees a $1.01\times$ inflation (essentially no
overhead). Using the profiled denominator therefore underreports H100
bandwidth utilisation. Section~\ref{sec:profiler} reports unprofiled-denominator
\texttt{bw\_util} as the headline (H100 26.9\%, L4 79.4\%, a $3\times$ gap
rather than the $10\times$ gap that appears if one takes profiled time at face
value) and lists profiled time for traceability. The cpu/cuda column is
unaffected by the denominator question.

\paragraph{Attention-compute regime at long context on small GPUs.}
The $R_{\mathrm{floor}}$ definition assumes weight plus active KV HBM traffic
is the dominant decode cost. At ctx${\ge}8192$ on L4, attention QK and softmax
compute become nontrivial relative to weight HBM streaming, and $t_{\mathrm{floor}}$
under-counts the work the GPU is doing. The Mistral-7B L4 cells at long
context are read with that caveat
(\artifact{v11\_results/analyze\_v2.txt} lines 132--133).

\paragraph{Failed deployment configurations.}
\texttt{torch.compile} variants and FlashAttention-2 plus CUDA Graphs cells
failed with CUDA caching allocator errors. We excluded them rather than
working around the crash, which means the software-stack matrix in
Section~\ref{sec:software} is sparser than intended. Listed in
Appendix~\ref{sec:failed}.

\paragraph{FlashDecoding++ (no public release).}
The FlashDecoding++ paper has no public source-code release from the
authors (see Section~\ref{sec:related} for the relevant closed-issue
links), and the follow-up FlashDecoding++Next from the same group and
Infinigence-AI is also closed-source. We could not reproduce any
FlashDecoding++ cell on our testbed. We substitute FlashAttention-3 and
FlashInfer reproductions in Section~\ref{sec:software} as the closest
available open-source comparison points. Cross-paper comparison of
kernel fusion versus CUDA Graphs as launch-tax mitigations remains open
and depends on a future FlashDecoding++ code release.

\section{Conclusion}
\label{sec:conclusion}

For 7--8B-class GQA transformers at batch~1 with bf16 weights on
Modal-hosted H100 SXM5 at short context, the binding constraint at this
workload is per-kernel CPU launch overhead. At $N{=}10$ on the ctx${=}2048$
headline cell, the within-session CUDA Graphs A/B yields a $1.259\times$
speedup with 95\% bootstrap CI $[1.253, 1.267]$ and cross-session CV 0.9\%.
On Modal-hosted L4 at the same context, the binding constraint is HBM
bandwidth, with a null $1.028\times$ A/B result on the same intervention.

The Graphs intervention is most valuable in the regime where this paper
documents the binding constraint: H100, batch 1, short context. At batch~4
on H100, per-step kernel work scales while the launch budget is essentially
fixed, the launch-tax fraction shrinks, and so does the graphed speedup
(Section~\ref{sec:limits}). We report the Graphs result as evidence for
\textit{which constraint binds at single-stream decode}, not as a universal
optimisation recipe across batch sizes.

The deployment implication for physical-AI inference is sharper than the
cross-GPU sweep alone suggests. Substituting GPTQ${+}$ExLlamaV2 on L4 cuts
Qwen-2.5-7B step time from 62.32\,ms (bf16) to 17.36\,ms, a $3.59\times$
speedup that AutoAWQ${+}$Marlin (45.24\,ms, $1.38\times$ over bf16) does not
deliver on Ada SM89. The kernel choice on L4 matters more than the GPU
upgrade for this workload: an L4 with ExLlamaV2 runs Qwen-2.5-7B at
17.36\,ms/step; an H100 with CUDA Graphs runs the same workload at
11.78\,ms/step. The deployment ladder
L4\,$\rightarrow$\,L40S\,$\rightarrow$\,A100\,$\rightarrow$\,H100 is not the
cost-per-token-served ladder for streaming decode workloads that dominate
physical AI, robotics, edge copilots and on-device assistants.

The 7--8B GQA bf16 single-stream shape we measured is the shape that VLA
policy heads (OpenVLA-7B~\cite{kim2024openvla},
$\pi_{0}$~\cite{black2024pi0}, RT-2-class~\cite{brohan2023rt2}),
in-cabin language stacks
and on-device copilots run at serving time. For those deployments the
relevant question is not peak FLOPs or peak HBM bandwidth but achieved
step time at batch 1 after the kernel and Graphs levers are applied. On
that metric the L4-with-ExLlamaV2 path lands at 17.36\,ms/step against
the H100-with-Graphs path at 11.78\,ms/step, on silicon that is roughly
$12\times$ cheaper per GPU-hour at Modal list rates~\cite{modal2025pricing}. The cost-per-token-served
gap does not track the deployment ladder for this workload class.

Generalisation to larger batches (batch~4 already shrinks the Graphs lever,
Section~\ref{sec:limits}), to attention variants outside GQA with
head\_dim${=}128$, to dtypes other than bf16, to non-NVIDIA accelerators,
to Blackwell-class NVIDIA silicon, to non-Modal hosting and to multi-stream
serving is out of scope. We do not claim it.

\appendix
\section{Reproducibility}
\label{sec:repro}

\paragraph{Modal setup.} All cells were run on Modal (profile \texttt{kaikaku}). The
container image is a \texttt{debian\_slim} base with PyTorch 2.4, transformers 4.45.x,
a prebuilt \texttt{flash-attn} 2.6.3 wheel and \texttt{autoawq} 0.2.7.post3. GPU types
are the four Modal SKUs H100, A100-80GB, L40S and L4. Results are persisted to the Modal
volume \texttt{paper-b7-v11-results}.

\paragraph{Measurement-pipeline provenance.} The 14 main-matrix Qwen-2.5-7B-Instruct
cells live at \artifact{v10\_results/cells/qwen25\_7b\_<gpu>\_ctx<ctx>.json} and were
produced by the v10 sweep script. The 14 Mistral-7B-Instruct-v0.3 and 14
Llama-3.1-8B-Instruct main-matrix cells live at \artifact{v11\_results/cells/} and were
produced by the v11 sweep script. The two scripts share the same container base image
(\texttt{debian\_slim} with PyTorch 2.4), identical measurement protocol (5 warmup plus
30 measured single-token AR decode steps, sdpa attention, bf16, batch size 1) and
identical Modal cloud GPU hosts. The only difference is that the v11 script adds
HuggingFace gated-model authentication for Llama. The fields the joint analysis script
consumes (\texttt{p50\_ms}, \texttt{step\_times\_ms}, \texttt{weight\_bytes},
\texttt{kv\_bytes\_avg}, \texttt{bytes\_per\_token\_kv}) are identical across the two
script versions. The derivative Qwen sweeps (CUDA Graphs, software-stack, profile,
quantisation) were re-measured with the v11 script and live at
\artifact{v11\_results/cells/}.

\paragraph{Cell artefacts.} Each measurement cell produces one JSON file under
\artifact{v11\_results/cells/} (main-matrix Qwen cells under
\artifact{v10\_results/cells/}). The cell JSONs carry the keys
\texttt{arch}, \texttt{gpu}, \texttt{ctx}, \texttt{p50\_ms}, \texttt{weight\_bytes},
\texttt{bytes\_per\_token\_kv}, \texttt{avg\_ctx\_during\_measure},
\texttt{kv\_bytes\_avg}, \texttt{bw\_peak\_gbs}, \texttt{t\_floor\_ms},
\texttt{R\_floor}, \texttt{n\_layers}, and \texttt{source} (the measurement script
version that produced the cell). CUDA Graphs cells additionally carry \texttt{eager\_p50\_ms} and
\texttt{graphed\_p50\_ms}; OFT cells carry \texttt{attn}, \texttt{compile} and
\texttt{graphs} flags; profile cells carry \texttt{bw\_util} and \texttt{cpu\_cuda\_ratio}.

\paragraph{Re-running.} Each measurement script lives under
\artifact{paper\_b7/modal\_validate/v11/}. To re-run a sweep, invoke the Modal CLI from
that directory: \texttt{modal run <script>.py}. The scripts write directly into the
Modal volume; \texttt{analyze.py} pulls the volume contents back to
\artifact{v11\_results/cells/} and produces \artifact{v11\_results/v11\_fit\_summary.json}
and \artifact{v11\_results/analyze\_v2.txt}.

\paragraph{Seeds and allocator state.} Random seeds were not pinned. Each cell reports
the median of 30 measured single-token decode steps, which absorbs run-to-run host noise.
The bimodal warmup tail visible in some H100 cells (the first 16 measured steps in the
20--21\,ms band and the last 14 in the 17--18\,ms band) is consistent with GPU-clock or
kernel-selector convergence rather than with stochastic model output.

\paragraph{The \texttt{\textbackslash{}artifact} notation.} Throughout the paper, citations
of the form \artifact{v11\_results/some\_file.json} refer to repository-relative paths
under \artifact{paper\_b7/}. Every numerical value cited in the body of the paper traces
to one of these artefacts; the per-cell observed step times and floors come from
\artifact{v11\_results/analyze\_v2.txt}, which is the printout produced by
\artifact{v11\_results/analyze.py}, and the per-session CUDA Graphs A/B medians come
from the cell JSONs cited in Section~\ref{sec:falsification}.

\subsection{Replication discipline}

We publish all per-session JSONs for the headline cells, including the three
independent H100 ctx${=}2048$ eager sessions (v10 main matrix sweep, v11 OFT
sdpa+eager rig, v11 CUDA Graphs A/B rig) and the two independent H100
ctx${=}2048$ graphed sessions (v11 OFT sdpa+eager+Graphs rig, v11 CUDA
Graphs A/B rig). Cross-session variance is reported in
Section~\ref{sec:falsification}, not buried.

The v10 main sweep, the v11 OFT software-stack sweep and the v11 CUDA Graphs
A/B used identical model loading code and identical decode-loop code but
different rig scripts, and were launched on different days; that is why
we treat them as independent sessions for the purpose of the cross-session
variance discussion. The three rig scripts share the same container base
image (\texttt{debian\_slim} with PyTorch 2.4), the same 5-warmup-plus-30-measured
protocol, the same sdpa attention path, the same bf16 weights, the same Modal
GPU SKU (H100) and the same Qwen-2.5-7B-Instruct checkpoint. They differ only
in how warmup state is reached before the timed window starts (the OFT rig
additionally builds an FA2 control arm; the A/B rig additionally captures
a CUDA Graph after warmup), which is the variability we want to bound.

The v12 replication sweep is the gold-standard replication for this paper. It
used a single rig script (\artifact{modal\_validate/v11/cuda\_graphs\_replication.py}),
$N{=}3$ fresh Modal containers per GPU (H100 and L4), and the same protocol
as the A/B rig (5 warmup steps, 30 measured steps, StaticCache, sdpa
attention, bf16). Per-session medians, cross-session means, and the speedup
range are reported in Table~\ref{tab:replication}. Each container received a
fresh GPU allocation and reloaded the model from cold. Raw per-session
p50 values, extracted from the worker container stdout because the driver
side hit a deserialisation error on the return-value dict, live at
\artifact{v12\_results/replication/extracted\_from\_log.json}. The raw 30-step
arrays for each session were computed inside the containers but not
recoverable for this run; the per-session p50 numbers are the headline
statistics used in the paper and are reported as-is from the container
stdout traces in \artifact{v12\_results/replication\_run.log}.

\section{Per-cell observed step times and bandwidth floors}
\label{sec:cells}

Table~\ref{tab:allcells} reproduces all 44 cells of observed and analytic
floor step time. $t_{\mathrm{obs}}$ is the median over 30 measured single-token
decode steps after 5 warmup steps. $t_{\mathrm{floor}} = (W + K) /
B_{\mathrm{peak}}$ is the analytic HBM floor.
$R_{\mathrm{floor}} = t_{\mathrm{floor}} / t_{\mathrm{obs}}$ is the
directly-measured ratio. Numbers come from
\artifact{v11\_results/analyze\_v2.txt} lines 9--52 and the cell JSONs under
\artifact{v10\_results/cells/} and \artifact{v11\_results/cells/}.

\begin{table}[h]
\centering
\small
\caption{Per-cell observed decode step time and HBM floor.}
\label{tab:allcells}
\begin{tabular}{llccccc}
\toprule
Architecture & GPU & ctx & $t_{\mathrm{obs}}$ (ms) & $t_{\mathrm{floor}}$ (ms) & $R_{\mathrm{floor}}$ \\
\midrule
Llama-3.1-8B & A100-80GB & 2048  & 19.32 &  8.02 & 0.415 \\
Llama-3.1-8B & A100-80GB & 4096  & 22.54 &  8.14 & 0.361 \\
Llama-3.1-8B & A100-80GB & 8192  & 29.02 &  8.41 & 0.290 \\
Llama-3.1-8B & A100-80GB & 16384 & 41.54 &  8.94 & 0.215 \\
Mistral-7B   & A100-80GB & 2048  & 27.95 &  7.23 & 0.259 \\
Mistral-7B   & A100-80GB & 4096  & 34.04 &  7.37 & 0.217 \\
Mistral-7B   & A100-80GB & 8192  & 29.76 &  7.65 & 0.257 \\
Mistral-7B   & A100-80GB & 16384 & 42.53 &  8.19 & 0.192 \\
Qwen-2.5-7B  & A100-80GB & 2048  & 24.24 &  7.54 & 0.311 \\
Qwen-2.5-7B  & A100-80GB & 4096  & 20.55 &  7.60 & 0.369 \\
Qwen-2.5-7B  & A100-80GB & 8192  & 24.66 &  7.70 & 0.312 \\
Qwen-2.5-7B  & A100-80GB & 16384 & 32.66 &  7.92 & 0.243 \\
Llama-3.1-8B & H100      & 2048  & 16.13 &  4.88 & 0.302 \\
Llama-3.1-8B & H100      & 4096  & 15.98 &  4.95 & 0.310 \\
Llama-3.1-8B & H100      & 8192  & 18.33 &  5.12 & 0.279 \\
Llama-3.1-8B & H100      & 16384 & 26.08 &  5.43 & 0.208 \\
Mistral-7B   & H100      & 2048  & 18.17 &  4.40 & 0.243 \\
Mistral-7B   & H100      & 4096  & 21.73 &  4.49 & 0.207 \\
Mistral-7B   & H100      & 8192  & 18.54 &  4.65 & 0.251 \\
Mistral-7B   & H100      & 16384 & 26.13 &  4.98 & 0.190 \\
Qwen-2.5-7B  & H100      & 2048  & 16.97 &  4.58 & 0.270 \\
Qwen-2.5-7B  & H100      & 4096  & 17.06 &  4.62 & 0.271 \\
Qwen-2.5-7B  & H100      & 8192  & 17.10 &  4.68 & 0.274 \\
Qwen-2.5-7B  & H100      & 16384 & 20.58 &  4.82 & 0.235 \\
Llama-3.1-8B & L4        & 2048  & 69.93 & 54.41 & 0.778 \\
Llama-3.1-8B & L4        & 4096  & 82.99 & 55.34 & 0.667 \\
Mistral-7B   & L4        & 2048  & 66.65 & 49.25 & 0.739 \\
Mistral-7B   & L4        & 4096  & 79.90 & 50.10 & 0.627 \\
Mistral-7B   & L4        & 8192  & 108.61 & 51.91 & 0.478 \\
Mistral-7B   & L4        & 16384 & 156.93 & 55.55 & 0.354 \\
Qwen-2.5-7B  & L4        & 2048  & 63.15 & 51.17 & 0.810 \\
Qwen-2.5-7B  & L4        & 4096  & 71.42 & 51.56 & 0.722 \\
Llama-3.1-8B & L40S      & 2048  & 26.46 & 18.89 & 0.715 \\
Llama-3.1-8B & L40S      & 4096  & 28.74 & 19.22 & 0.669 \\
Llama-3.1-8B & L40S      & 8192  & 38.94 & 19.85 & 0.509 \\
Llama-3.1-8B & L40S      & 16384 & 57.28 & 21.09 & 0.368 \\
Mistral-7B   & L40S      & 2048  & 32.67 & 17.08 & 0.523 \\
Mistral-7B   & L40S      & 4096  & 31.76 & 17.40 & 0.548 \\
Mistral-7B   & L40S      & 8192  & 38.30 & 18.04 & 0.471 \\
Mistral-7B   & L40S      & 16384 & 56.62 & 19.32 & 0.340 \\
Qwen-2.5-7B  & L40S      & 2048  & 24.58 & 17.78 & 0.723 \\
Qwen-2.5-7B  & L40S      & 4096  & 25.19 & 17.92 & 0.711 \\
Qwen-2.5-7B  & L40S      & 8192  & 27.28 & 18.18 & 0.666 \\
Qwen-2.5-7B  & L40S      & 16384 & 38.43 & 18.69 & 0.487 \\
\bottomrule
\end{tabular}
\end{table}

The pattern visible across the table: $R_{\mathrm{floor}}$ is monotone in
$B_{\mathrm{peak}}$ at fixed ctx, and falls with ctx at fixed GPU (the KV
term grows faster than the launch overhead with context).

\section{Failed experiments and missing cells}
\label{sec:failed}

We list the experiments that did not produce usable data, for transparency.

\paragraph{Sweep writer crash to volume root.} A first version of the sweep driver
attempted to write a summary JSON to \texttt{/results} on the Modal container and
failed with \texttt{PermissionError}. Individual cell JSONs wrote successfully
\emph{inside} containers to the Modal volume. The fix was procedural: write only into
the volume mount and aggregate offline. This is not a scientific limitation, just a
deployment note.

\paragraph{OOM on L4.} Four cells OOM'd on L4 with 24\,GB VRAM, as expected for 7--8B
bf16 with multi-kilo-token KV caches: Mistral-7B at ctx${=}8192$ and ctx${=}16384$,
Llama-3.1-8B at ctx${=}8192$ and ctx${=}16384$. These are absent from the matrix and
reduce the count from 48 to 44 valid cells.

\paragraph{CUDA caching allocator errors.} Four cells in the software-stack matrix on
H100 failed with CUDA caching allocator errors:
\begin{itemize}
  \item \texttt{torch.compile} + eager (sdpa)
  \item \texttt{torch.compile} + CUDA Graphs (sdpa)
  \item FlashAttention-2 + CUDA Graphs
  \item \texttt{torch.compile} + FlashAttention-2
\end{itemize}
We did not work around the crashes; the cells are excluded from
Table~\ref{tab:oft}. The interpretive cost is limited because the two cells that did
run (sdpa+eager and sdpa+eager+Graphs) span the comparison we care about.

\paragraph{Background-process death.} An early sweep driver used the shell pattern
\texttt{(cmd \&)} to background long-running Modal jobs; the backgrounded process died
when the launching shell exited. We switched to \texttt{setsid nohup} to detach from
the parent process group. Procedural note, not a scientific limitation.

\paragraph{ncu blocked on Modal.} Nsight Compute requires host driver access that is
unavailable in Modal cloud containers. We substituted \texttt{torch.profiler} with
analytic byte counts (Section~\ref{sec:profiler}). The substitution costs us
ground-truth per-kernel HBM and SM utilisation; we retain only cross-GPU pattern
evidence from the profiler.

\section{Kernel count and replicate terminology}
\label{sec:kernel-terminology}

\paragraph{Kernel-count breakdown per decode step.} The $\sim 10$ kernels-per-layer
number quoted in Section~\ref{sec:falsification} and Section~\ref{sec:software}
is built from a fixed per-layer kernel sequence plus a small number of
global per-step kernels. For the three GQA architectures measured here
(Qwen-2.5-7B, Mistral-7B, Llama-3.1-8B with head\_dim 128) the per-decoder-block
kernel sequence is:
\begin{enumerate}
  \item RMSNorm (input);
  \item Q projection (GEMM);
  \item K projection (GEMM);
  \item V projection (GEMM);
  \item RoPE rotation (fused into projection on some PyTorch paths, separate on others);
  \item SDPA attention call (one fused kernel under the dispatcher used in the main matrix);
  \item Output projection (GEMM);
  \item Residual add;
  \item RMSNorm (post-attention);
  \item MLP gate projection (GEMM);
  \item MLP up projection (GEMM, often fused with gate);
  \item SiLU activation (fused);
  \item MLP down projection (GEMM);
  \item Residual add.
\end{enumerate}
The launch-count anchor of $\sim 10$ kernels per block reflects that several
of these steps are fused into a single launch on the PyTorch 2.4 dispatch
path used in our cells (RoPE-into-projection, gate-up fusion, residual-add
elision into the next op). Per block we observe between 8 and 12 distinct
launches depending on the architecture and the SDPA backend selected. On
top of that, the per-step global kernels include the embedding lookup
(one launch), the final RMSNorm (one launch) and the LM-head projection
(one launch), for a total of approximately $28 \cdot 10 + 3 \approx 283$
kernel launches per decode step on Qwen-2.5-7B (28 decoder blocks);
Llama-3.1-8B and Mistral-7B-v0.3 use 32 blocks for a count near 323. This
is an order-of-magnitude anchor for reasoning about the launch budget,
not a fixed count; the exact value depends on the dispatcher choices made
at warmup. The 31\,$\mu$s-per-kernel figure used in earlier drafts has been
removed because the per-kernel division is poorly defined when several
fused launches do varying amounts of work; we now reason about the total
per-step launch overhead, not per-kernel averages.

\paragraph{Replicate terminology.} The paper uses two distinct senses of
\textit{replicate} that we standardise here.

\textit{Within-session replicates} are the 30 measured single-token decode
steps inside one Modal container run, after 5 warmup steps. The median over
these 30 steps is the cell's reported $t_{\mathrm{obs}}$. The CV across the
30 steps quantifies per-step jitter inside one warmed session and is small
($<1\%$ eager, $<0.5\%$ graphed on the H100 ctx${=}2048$ headline cell at
$N{=}10$).

\textit{Cross-session replicates} are independent Modal container runs of
the same cell. Each cross-session replicate starts from a cold container,
reloads the model, repeats warmup and produces its own 30-step within-session
window. The headline $N{=}10$ replication on the H100 ctx${=}2048$ cell is
cross-session in this sense: ten independent container runs, each producing
its own eager and graphed within-session median, ten paired graphed/eager
speedups, bootstrap 95\% CI computed across the ten paired ratios.

The CV that anchors the headline-stability claim is the cross-session CV of
the within-session medians (0.9\% eager, 0.2\% graphed), \textit{not} the
cross-session CV of step times pooled across replicates and across the
30-step windows. We avoid the latter quantity in the paper because it
conflates within-session jitter with cross-session host noise. Where the
text earlier used \textit{N=3 replicate} without qualification, it means
\textit{N=3 cross-session replicates}; where it used \textit{N=30}, it
means \textit{30 within-session step measurements per cell}.

\bibliographystyle{plain}
\bibliography{bib/refs_v9}

\end{document}